# Observation of Negative THz Photoconductivity in Large Area Type-II Dirac Semimetal $PtTe_2$

Peng Suo[1†], Huiyun Zhang[2†], Shengnan Yan[3†], Wenjie Zhang[1], Jibo Fu[1], Xian Lin[1], Song Hao[3], Zuanming Jin[5,6], Yuping Zhang[2], Chao Zhang[4], Feng Miao[3], Shi-Jun Liang[3*], and Guohong Ma[1,6*]

[1]Department of Physics, Shanghai University, Shanghai 200444, China

[2]College of Electronics and Information Engineering, Shandong University of Science and Technology, Qingdao 266590, China

[3]National Laboratory of Solid State Microstructures, School of Physics, Collaborative Innovation Center of Advanced Microstructures, Nanjing University, Nanjing, 210093, China

[4]School of Physics, University of Wollongong, Wollongong, New South Wales 2522, Australia

[5]Terahertz Technology Innovation Research Institute, Terahertz Spectrum and Imaging Technology Cooperative Innovation Center, Shanghai Key Lab of Modern Optical System, University of Shanghai for Science and Technology, 516 JunGong Road, Shanghai 200093, China

[6]STU & SIOM Joint Laboratory for Superintense Lasers and the Applications, Shanghai 201210, China

[†]These authors contributed equally

[*]To whom correspondence should be addressed, email address: sjliang@nju.edu.cn (S.-J. Liang) and ghma@staff.shu.edu.cn (G. H. Ma)

ABSTRACT

As a newly emergent type-II Dirac semimetal, Platinum Telluride ($PtTe_2$) stands out from other 2D noble-transition-metal dichalcogenides for the unique band structure and novel physical properties, and has been studied extensively. However, the ultrafast response of low energy quasiparticle excitation in terahertz (THz) frequency remains nearly unexplored yet. Herein we employ optical pump-terahertz (THz) probe spectroscopy (OPTP) to systematically study the photocarrier dynamics of $PtTe_2$ thin films with varying pump fluence, temperature, and film thickness. Upon photoexcitation the THz photoconductivity (PC) of $PtTe_2$ films show abrupt increase initially, while the THz PC changes into negative value in a subpicosecond time scale, followed by a prolonged recovery process that lasted a few ns. The magnitude of both positive and negative THz PC response shows strongly pump fluence dependence. We assign the unusual negative THz PC to the formation of small polaron due to the strong electron-phonon (e-ph) coupling, which is further substantiated by temperature and film thickness dependent measurements. Moreover, our investigations give a subpicosecond time scale of sequential carrier cooling and polaron formation. The present study provides deep insights into the underlying dynamics evolution mechanisms of photocarrier in type-II Dirac semimetal upon photoexcitation, which is crucial importance for designing $PtTe_2$-based optoelectronic devices.

## INTRODUCTION

The emergence of three dimensional (3D) Dirac semimetals (DSMs)，a 3D analog of graphene, has lately captured immense attention because of their nontrivial topology properties such as exotic Fermi arc surface states, anomalous giant magnetoresistance, and topological nontrivial quantum oscillation, and the potential applications in optoelectronic devices [1-5]. Recently, the Lorentz-violating type-II Dirac semimetals (*e.g.* $PdTe_2$, $PtTe_2$ and $PtSe_2$) with tilted Dirac cones in certain momentum direction have been predicted theoretically and confirmed experimentally[6-10], in which the anisotropic and heavily titled bulk Dirac cones are formed by two crossed valence bands [6,11-14]. Conventionally, these DSMs films show high electrical mobility, and the film conductivity is strongly relevant to the film thickness as well as the temperature [15,16]. Benefiting from the linear dispersion band structure, ultrahigh carrier mobility and zero band-gap, DSMs show a huge potential toward a high-performance photodetector with high operation speed, broadband response from ultraviolet to THz [17,18]. Over past few years, photoconductive response in DSMs has been studied widely in order to uncover the photocarrier scattering mechanism governed by the unique gapless band structure protected by crystalline symmetry [17,19,20]. In conventional semiconductor and semimetal, photoexcitation across bandgap generally generates positive photoconductivity (PPC) from direct current (DC) to THz frequency. In addition, the observations of negative photoconductivity (NPC) in THz frequency have been also widely reported in some materials such as graphene, TMDs and topological insulators, which were proposed to arise from the role of hot electrons, formation of trions and topological surface states, respectively [21-24]. It is also noted that the THz PC in graphene can be positive or negative depending upon the Fermi energy, doping behavior and environmental gases [22,25-28]. In contrast, $Cd_3As_2$, a type-I DSM, always shows THz PPC [29,30]. Due to the different band dispersion, the THz PC response in type-II DSMs can be very different from that of type-I counterpart. As a model material of the type-II DSMs, the band structure and transport properties in $PtTe_2$ have been investigated extensively [31-34]. As far as we know, there has no report about the THz PC study in the type-II DSMs so far, and the THz photoconductive response and associated underlying mechanism remain elusive, although the recent reports demonstrate the tremendous application potential of the DSM $PtTe_2$ in long-wave detection and optoelectronic devices [35,36].

In the present work, we employed OPTP spectroscopy to unravel the ultrafast photocarrier dynamics of type-II Dirac semimetal $PtTe_2$ thin film in ultrafast time scale. Our experimental results reveal that the photocarrier responses of $PtTe_2$ thin films initially exhibits absorption enhancement of THz radiation induced by hot electron, subsequently the enhanced THz transmission takes place within a subpicosecond time scale, which is rooted in the formation of small polaron induced reduction of carrier mobility. Long recovery process is dominated by the dissociation of small polaron. To the best of our knowledge, this is the first observation of small polaron formation in $PtTe_2$, and also the first detailed study about the photocarrier dynamics of type-II DSMs in nonequilibrium state, which might pave the new way for the design of electronic and optoelectronic devices.

# EXPERIMENTAL DETAILS

**Sample characterization.** The high quality films $PtTe_2$ without intended doping were synthesized on the fused silica substrate with the thickness of 6.8, 20 and 44 nm, and the detailed process have been provided in our prior work [16]. The detailed thickness and X-ray photoemission spectra of the three films are given in Figs. S1 and S2. Figures 1(a) and 1(b) display the crystal structure of $PtTe_2$ from both the top and side views. The background carrier densities at room temperature are in order of $10^{22}$ $cm^{-3}$ obtained from Hall measurement for the 6.8, 20 and 44 nm $PtTe_2$ film, respectively [Fig. S3 in SM], all films show metal-like property. Figure 1(c) shows the Raman spectrum of the $PtTe_2$ film with excitation laser line of 532 nm, two pronounced phonon modes at 112 $cm^{-1}$ ($E_g$) and 156 $cm^{-1}$ ($A_{1g}$) are clearly seen for all films, which correspond to the in-plane and out-of-plane vibration mode, respectively. Interestingly, the intensity ratio of $E_g$ and $A_{1g}$ modes ($E_g/A_{1g}$) increases with film thickness, which suggests the in-plane vibration is more sensitive than out-of-plane vibration to the film thickness. As displayed in Figure 1(d), the room temperature X-ray diffraction (XRD) pattern measured with grazing incidence exhibits pronounced (001) characteristic peak for all films, and two weak peaks indexed as (011) and (012) are observed clearly for 15 nm $PtTe_2$ film, indicating good crystallinity of our samples.

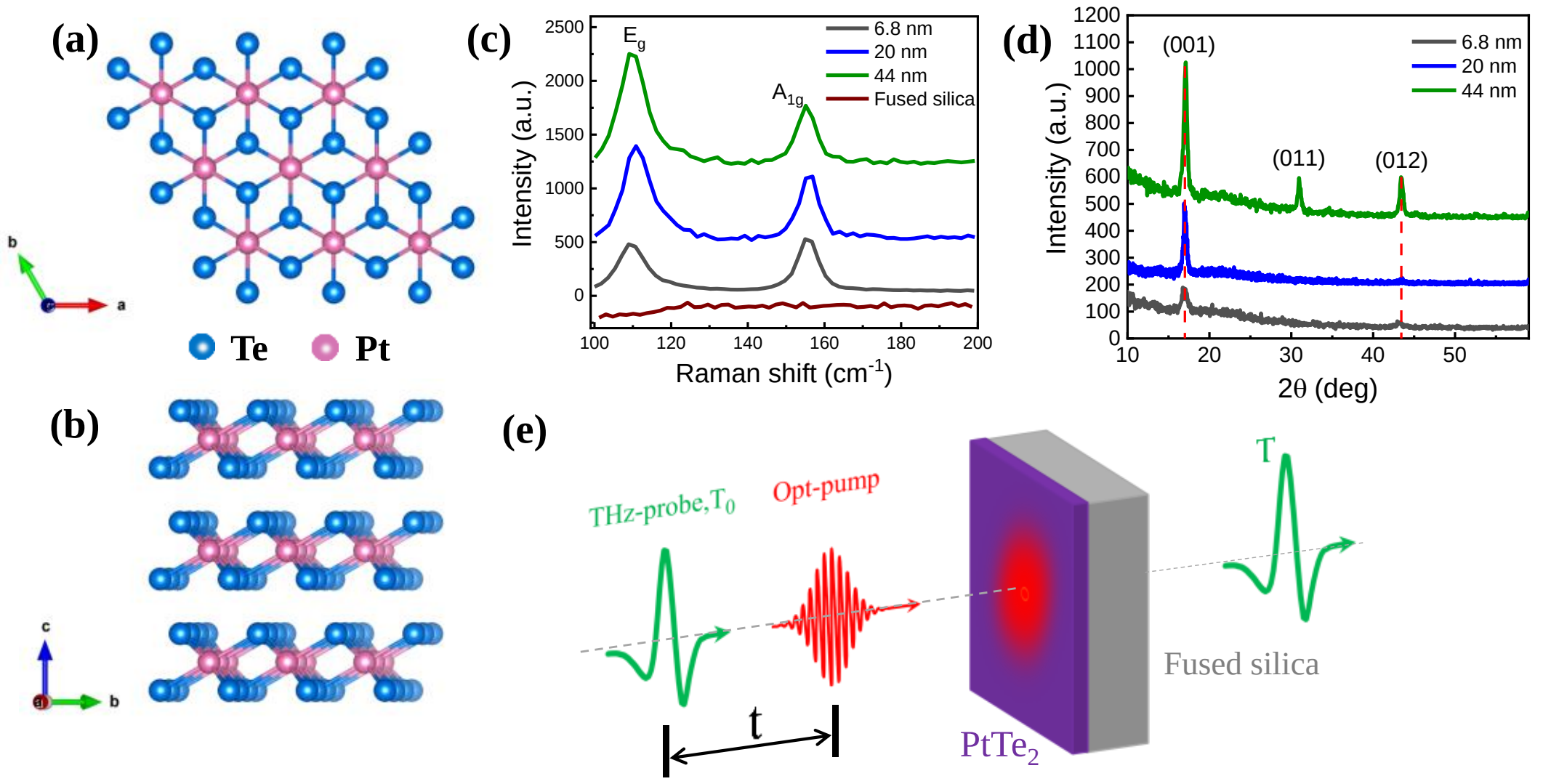

**Figure 1.** (a) and (b) show the crystal structure of *ab* and *bc* side view, respectively. (c) Raman spectra and (d) X-ray diffraction pattern of $PtTe_2$ films. (e) Schematic diagram of optical pump-THz

probe spectroscopy.

**Transient THz dynamics measurement.** The time-resolved OPTP experiments in the transmission configuration were performed to explore the dynamics of photocarriers. The optical pulses are delivered from a Ti: sapphire amplifier with 120 femtoseconds (fs) duration at central wavelength of 780 nm (1.59 eV) and a repetition rate of 1 kHz. The THz emitter and detector are based on a pair of (110)-oriented ZnTe crystals. The optical pump and THz probe pulse are collinearly polarized with a spot size of 6.5 and 2.0 mm on the surface of sample, respectively. All measurements were conducted in dry nitrogen atmosphere to avoid the absorption of water vapor, and the sample was placed in a cryostat with temperature varying from 5 K to 300 K.

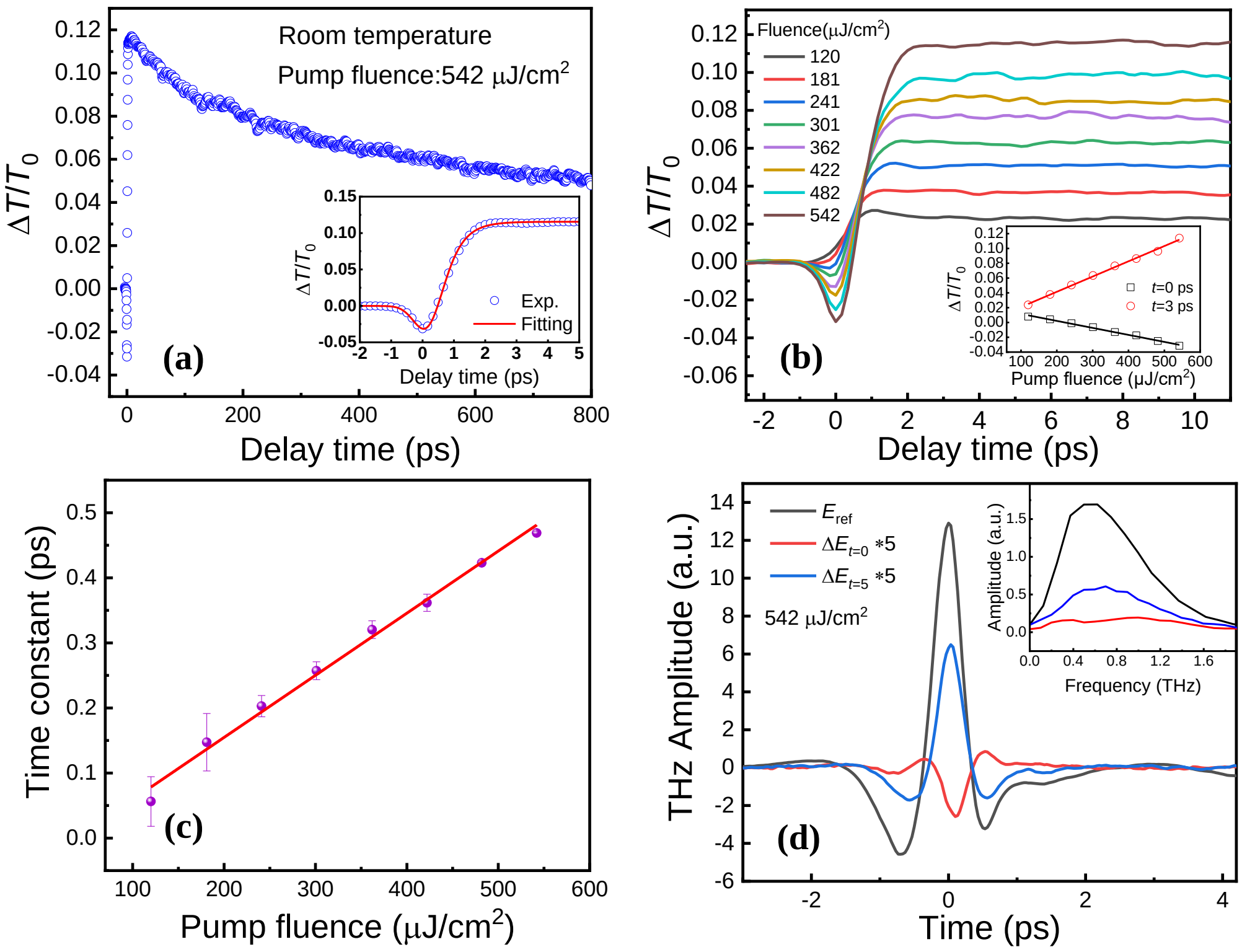

**Figure 2.** (a) The transient dynamics of 20 nm $PtTe_2$ film under 780 nm pump at room temperature. The inset plots the transient response in a window of short-scan time, and the red line represents monoexponential fitting curve. (b) The THz transmission, $\Delta T/T_0$, as a function of pump-probe delay time under various pump fluences. The inset plots the magnitude of $\Delta T/T_0$ at $t$=0 ps (black) and $t$=3 ps (red) with respect to the pump fluence, respectively. (c) The rising time constants obtained from monoexponential fitting with respect to pump fluence, and the solid line is a liner fitting. (d) THz

waveform transmission of the sample without pump (black) as well as with pump fluence of 542 μJ/cm$^2$ collected at delay time $t$=0 (red) and $t$=5 ps (blue), respectively. The inset shows the corresponding frequency spectra obtained from Fourier transform.

## EXPERIMENTAL RESULTS

Figure 1(e) illustrates the schematic diagram for OPTP spectroscopy. To probe the photocarrier relaxation of $PtTe_2$ films, we have applied THz pulse after an optical pump at 780 nm and measured the photoinduced transmission change, which is defined as $\Delta T=T-T_0$ with respect to the delay time, $t$, between THz pulse and optical pulse, where the $T_0$ and $T$ denote the transmission signals of THz electric field peak value without and with photoexcitation, respectively. In a thin film approximation, the pump induced the THz transmission change $\Delta T/T_0$, is proportional to the THz NPC $-\Delta\sigma$, i.e. $\Delta T/T_0 \propto -\Delta\sigma$. In the following, the term "film" refers to the 20-nm $PtTe_2$ film on fused silica substrate unless otherwise stated.

Figure 2 illustrates the pump fluence dependent transient THz response of the film at room temperature. Figure 2(a) shows the photoinduced transient trace $\Delta T/T_0$ as a function of delay time with pump fluence of 542 μJ/cm$^2$. The inset plots the transient THz response in a window of short-scan time. Fascinatingly, after photoexcitation, the transient THz transmission consists of three stages with distinct time scale: (i) a pump-induced rapid increase in THz PC (drop in THz amplitude transmission) with response time limited by the laser pulse duration; (ii) the rapid transition in THz PC signal from positive to negative occurring on subpicosecond time scale; (iii) the slow recovery process lasts a few *ns* from the maximum bleach signal to equilibrium state. Notably, the transient THz response after photoexcitation in the film shows clear difference from that in type-I DSMs, such as graphene and $Cd_3As_2$, in which either positive or negative THz PC responses were observed [21,22,29,30]. It should be mentioned that fused silica substrates show neglected THz response under the identical photoexcitation. We would like to stress that we have also fabricated identical $PtTe_2$ films on sapphire and yttrium aluminum garnet (YAG) substrates. All these films show similar THz response as demonstrated in note 4 and 5 of SM.

Figure 2(b) shows the transient THz transmission traces under various pump fluences,

and the inset presents the magnitude of $\Delta T/T_0$ with respect to the pump fluence at delay time of $t$=0 (black) and $t$=3 ps (red), respectively. The good linear fluence dependent THz transmission indicates that the transient THz responses do not exhibit saturating behavior up to pump fluence of 542 μJ/cm$^2$. Considering that the bleaching signal last a few *ns*, we employ a monoexponential function convoluted with laser pulse to fit the ultrafast transient THz transmission response of Fig 2(b) in the initial 10 ps time window, with the fitting results shown in Figure S6 of SM. The fitting time constant as a function of pump fluence are presented in Figure 2(c). The result indicates that the relaxation time of THz PC from positive to negative increases linearly with the applied pump fluence. Figure 2(d) shows the transmitted THz waveforms through the film at $t$=0 ps (red), 5 ps (blue) and without photoexcitation (black), respectively. The photoinduced change in the THz electric field is expressed as $\Delta E(t_s, t)=E(t_s, t)-E_{\mathrm{ref}}(t_s)$, where $E(t_s, t)$ and $E_{\mathrm{ref}}(t_s)$ is the time-domain THz waveform at $t$ with and without pump, respectively, and $t_s$ is the electro-optical sampling delay for the distinction from the pump-probe delay time $t$. It is clear that THz transmitted waveforms collected at $t$=0 ps and $t$=5 ps show out-of-phase with each other, indicating that the photobleaching of THz transmission at $t$=5 ps occurs after photoexcitation. The relevant frequency-domain spectra via Fourier transform of $E_{\mathrm{ref}}$, $\Delta E_{t=0}$ and $\Delta E_{t=5}$ are showed in the inset of Figure 2(d).

The PC of the film, $\Delta\sigma=\Delta(n\mu)e$ with $e$ being elementary charge, is determined by the change of carrier concentration ($\Delta n$) and mobility ($\Delta\mu$) induced by photoexcitation. Thus, the time evolution of THz transmission response after photoexcitation is governed by $\Delta n$ and $\Delta\mu$ [21,37]. Compared to the intrinsic high carrier concentration (in the order of $10^{22}$ cm$^{-3}$) of the film under studied, the photogenerated carrier density (~$1.8\times10^{20}$ cm$^{-3}$ for highest pump fluence) is less than 1% and negligible. The photoexcitation mainly results in the thermalization of electrons through electron-electron (e-e) scattering within time scale of several tens of fs. The resulting high electron temperature would broaden the Fermi distribution of hot carriers over a wider energy range and lead to free carriers' absorption of THz pulses due to the intraband transition undergoes larger possible momentum and energy conservation spaces [30].

Therefore, the photoexcitation generated hot carriers give rise to the sharp enhancement of THz absorption in the film. Considering the fact that the bleaching signal appears after a fast relaxation of subpicosecond time scale, the THz NPC should only arise from the photoexcitation-induced reduction in carrier mobility of the film. Next we will discuss origins that may lead to the diminution in carrier mobility of the film.

## DISCUSSIONS

Before assigning the THz NPC to the excitation of small polaron in $PtTe_2$, we discuss other possible origins of the NPC, with more detailed interpretations provided in the note 7 in SM. In brief, one possible origin of the observed NPC in $PtTe_2$ film could be due to the increased electron effective mass after photoexcitation, which would be true if the electrons in Dirac cones are photoexcited into conduction band under 1.59-eV-optical-pump as used in our experiments. However, it is not the case since the Dirac point in $PtTe_2$ is located around 0.8 eV below the Fermi surface according to the literatures [6,7]. We have carried out OPTP measurement with pump wavelength of 1600 nm (0.75 eV), the pump-induced bleaching of THz transmission is still clearly observed as shown in Fig. S7.1 in SM. Therefore, the conjecture about the contribution of Dirac electrons can be easily excluded. Another possibility is that the observed NPC in $PtTe_2$ arises from the formation of trion after photoexcitation as in monolayer $MoS_2$ [23]. Considering the metallic nature of the film with high carrier concentration, the formation of the exciton with photoexcitation can be totally screened by the background free carriers, thus this possibility can be safely ruled out. Thirdly, photoexcitation leading to the elevated lattice temperature may be possible origin of the observed NPC in $PtTe_2$ film. According to two-temperature model, the e-e thermalization is much faster than e-ph thermalization after photoexcitation, and the electrons with the elevated temperature transfer the excess energy to lattice via e-ph coupling until the two subsystems reach a balanced temperature [38,39]. The electronic temperature cannot be lower than that before photoexcitation, therefore it is impossible that the observed NPC comes from hot carrier contribution. In addition, $PtTe_2$ films with identical thickness of 20 nm were grown on various substrates with different thermal conductivities (fused silica, YAG and sapphire), the recovery processes of transient THz signals (Fig. S7.2 in SM) show almost same relaxation behaviors, which suggests that thermal diffusion cannot explain the long-lived relaxation. Fourthly, impurity in the $PtTe_2$ film may also lead to the NPC after photoexcitation as observed in some semiconductors [40-42]. We exclude the contribution of impurity to the NPC based on the following two facts: one is that the THz conductivity (without pump) increases with decreasing temperature as shown in Fig. S7.3 of SM, which indicates the metallic nature of the $PtTe_2$ films and the negligible impurity contribution to THz conductivity. The other is that the relaxation time (*i.e.* a few *ns*) of the NPC observed in Figure 2(a) is 4~6 orders of magnitude faster than the recovering time of the NPC reported in semiconductors [40,41,43,44]. Last but not least, we mention that THz radiation after photoexcitation on the surface of $PtTe_2$ is negligible so that the observed NPC signal can also rule out the contribution from the photo-induced THz radiation. Based on the analyses above, we interpret the unexpected NPC as the photo excitation of small polaron in $PtTe_2$, in which the formation of small polaron leads to the significant reduction of carrier mobility due to the phonon "dressing" of carrier.

The polaron has been experimentally observed in numerous compounds including metal oxides [45,46] and organic semiconductors [47] as well as halide perovskites [48-50]. Besides, the possible existence of polaron formation in graphene and TMDs has been proposed theoretically [51-53]. The strong and short-range e-ph coupling could give rise to the formation of small polaron. Previous experimental measurement [54] has shown that $PtTe_2$ exhibits a strong e-ph interaction with the coupling constant ranging from 0.38 to 0.42. Upon photoexcitation, the photogenerated carriers would couple with $PtTe_2$ lattice vibration strongly. It is this strong e-ph coupling that makes the cooling of hot carriers and deformation of the lattice around the carriers and gives rise to small polaron with reduced carrier mobility, which is the origin of the THz NPC. The polaron formation after photoexcitation is supported by the experimental results about the pump fluence dependent NPC shown in Figure 2(b). Higher pump fluence leads to a higher electron temperature in $PtTe_2$ film, therefore more phonon modes are excited during the hot electron cooling process via e-ph scattering due to the fact that the excited phonon modes are proportional to the temperature difference between the hot electron and the cold lattice. As a result, more electrons are "trapped" by phonons and the NPC signal under higher pump fluence becomes more pronounced. We also note that the rising time in Figure 2(c), *i.e.* the time scale for polaron formation, is within 100-500 fs and is consistent with ~100 fs reported in the conjugated polymer [55] and ~ 400 fs in the lead-iodide perovskites [48,49].

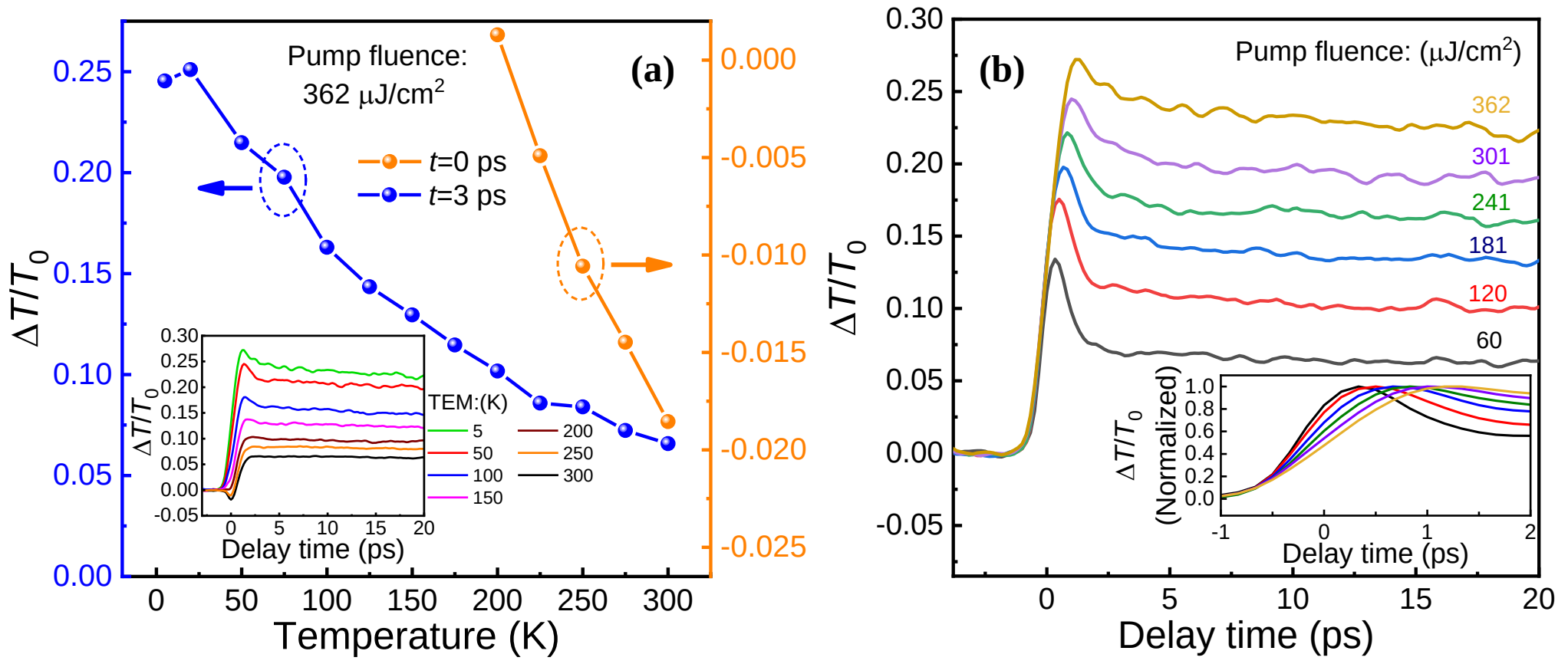

**Figure 3.** (a) Temperature dependence of $\Delta T/T_0$ at delay time of *t*=0 ps (orange) and 3 ps (blue), respectively. Inset shows the transient THz transmission at various temperatures at fixed pump fluence of 362 μJ/cm$^2$. (b) The transient THz transmission at temperature of 5 K under various pump fluence. Inset shows the enlargement of the rising process of $\Delta T/T_0$ under various pump fluences.

Since the mobility of small polaron decreases with the reduction of temperature [56], we anticipate that a more prominent NPC phenomenon occurs when reducing the temperature of the film. Figure 3(a) plots the $\Delta T/T_0$ collected at *t*=0 and 3 ps with respect to temperature, respectively, and the inset shows the temperature dependent temporal THz transmission at fixed pump fluence of 362 μJ/cm$^2$. It is clear that the NPC signal at *t*=3 ps increases monotonously with decreasing temperature, while the PPC signal around *t*=0 ps is present only at high temperature. When temperature is lower than 200 K, the PPC behavior vanishes completely. On the one hand, this phenomenon that the NPC signal is dramatically enhanced with decreasing temperature is consistent with the transport property of small polaron that the mobility decreases with decreasing temperature. On the other hand, the disappearance of PPC signal around *t*=0 ps at low temperature suggests that photocarrier thermalization and polaron formation take place simultaneously due to the significant increase of e-ph coupling strength at lower temperature. As a result, the newly formed polaron is "hot", which is composed of hot photocarrier and cold lattice, and subsequent polaron cooling process occurs via e-ph thermalization by transferring the excess energy to deformed lattice. Figure 3(b) shows the transient THz dynamics at 5 K for various pump fluences (transient THz dynamics at 100 K, 200 K and 300 K are shown in Fig. S8 of SM). Apparently, the relaxation of the NPC signal shows additional fast process with a typical time constant of ~1 ps by comparing with the signal measured at room temperature. The fast relaxation component represents the polaron cooling process, resulting in the increase of polaron photoconductivity. The inset in Fig. 3(b) shows zoom-in view of the rising process of the THz transmission, and it is seen that the rising time increases with the pump fluence. The slower rising time under higher pump fluence manifests itself in cooling of more hot electrons during the polaron formation, as a result, the cooling process of the newly formed polaron is not as pronounced as that under low pump fluence. Hence, the temperature dependent transient THz transmission spectra do support our interpretation that the NPC after photoexcitation in $PtTe_2$ films comes from the formation of small polaron.

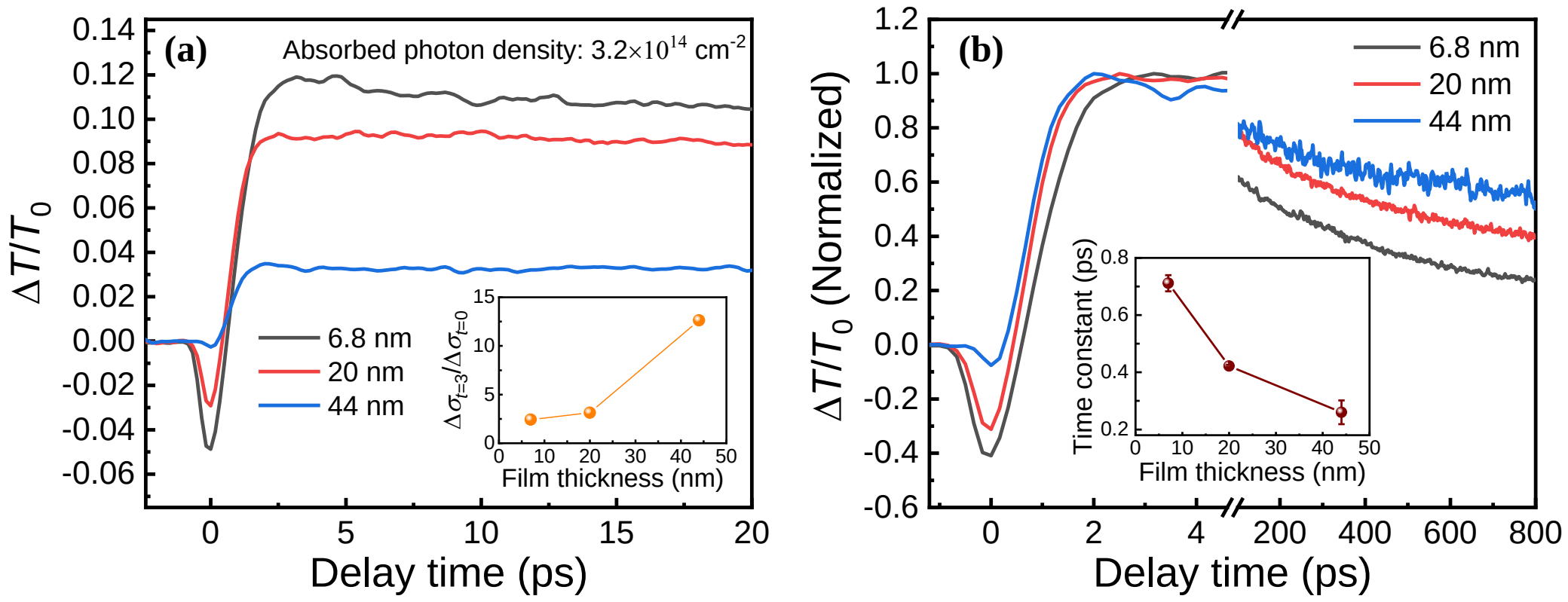

**Figure 4.** (a) The transient THz transmission for 6.8, 20 and 44 nm $PtTe_2$ films under 780 nm pump at room temperature, the optically injected photon density is $3.2\times10^{14}$ $cm^{-2}$ in all films. Inset plots the magnitude ratio of $\Delta\sigma_{t=3}/\Delta\sigma_{t=0}$ with respect to the film thickness. (b) The normalized THz transmission in long scan window for the three films, inset plots the fitting rise time with respect to the film thickness, the solid line is a guide to the eyes.

We further study the effect of film thickness on the THz PC short- and long-time windows at room temperature, with the results shown in Figure 4. The inset in Figure 4(a) shows the magnitude ratio of NPC collected at $t$=3 ps and PPC at $t$=0 ps, increase of the ratio with the film thickness shows that thicker film exhibits more pronounced negative THz PC. Figure 4(b) shows the normalized transient THz transmission in a long scan window for the three thickness films, longer relaxation time present in the thicker film indicates slower polaron dissociation process in thicker film with stronger e-ph coupling. Inset in Figure 4(b) presents the fitted rising time with respect to film thickness (see Fig. S9 in SM for the fitting details). Faster rising time occurring in the thicker film is indicative of larger e-ph coupling in thicker $PtTe_2$ film again, which is reasonable that thicker film has better crystalline as supported by the Raman and XRD results shown in Fig. 1 (c) and (d), respectively.

We have evaluated carrier mobility as well as the pump induced mobility change of 20 nm $PtTe_2$ film in THz frequency. The average THz conductivity $\sigma_0$ obtained from THz time domain spectrum is ~$3.3\times10^5$ S $m^{-1}$, and the carrier mobility $\mu_0$ is calculated to be 1.5 $cm^2$ $V^{-1}$ $s^{-1}$, which is close with previous report [15] (Note 10 in SM presents the detailed extraction). The pump induced carrier mobility change is calculated to be $\Delta\mu/\mu_0=(\mu-\mu_0)/\mu_0=10.5\%$ at delay time of 3 ps under the absorbed photon density of $3.2\times10^{14}$ $cm^{-2}$. The dramatically reduced carrier mobility at high pump fluence and low temperature (Fig. S10.2 in SM) agrees with our model of small polaron formation stemming from the photogenerated hot carrier induced lattice deformations in $PtTe_2$ films.

## CONCLUSIONS

To summarize, we have utilized ultrafast optical pump-THz probe spectroscopy to investigate the relaxation dynamics of photoexcited carriers in type-II Dirac semimetal $PtTe_2$ films. The significant feature of the transient THz transmission is the enhanced THz absorption followed by THz photobleaching signal with lifetime of a few ns. The fast THz response with time constant of ~1.0 ps is dominated by sequential carriers cooling and formation of small polaron, and the dissociation of small polaron plays a leading role in recovery process of slow THz bleaching. Our investigations offer insights into the photocarriers' dynamics in type-II Dirac semimetal and the new findings may find their implications in the development of novel optoelectronic devices.

## ACKNOWLEDGMENTS

This work is supported by the National Natural Science Foundation of China (NSFC, Nos. 11674213, 61735010, 11874264, 61875106, 61775123, 61974176, 61625402). Science and Technology Commission of Shanghai Municipality (Shanghai Rising-Star Program 18QA1401700), Natural Science Foundation of Jiangsu Province (BK20180330).

## REFERENCES:

[1] S.-Y. Xu, C. Liu, S. K. Kushwaha, R. Sankar, J. W. Krizan, I. Belopolski, M. Neupane, G. Bian, N. Alidoust, T.-R. Chang, H.-T. Jeng, C.-Y. Huang, W.-F. Tsai, H. Lin, P. P. Shibayev, F.-C. Chou, R. J. Cava, and M. Z. Hasan, Science **347**, 294 (2014).

[2] Z. K. Liu, B. Zhou, Y. Zhang, Z. J. Wang, H. M. Weng, D. Prabhakaran, S.-K. Mo, Z. X. Shen, Z. Fang, X. Dai, Z. Hussain, and Y. L. Chen, Science **343**,864 (2014).

[3] S. Borisenko, Q. Gibson, D. Evtushinsky, V. Zabolotnyy, B. Buchner, and R. J. Cava, Phys. Rev. Lett. **113**, 027603 (2014).

[4] F. Fei, X. Bo, R. Wang, B. Wu, J. Jiang, D. Fu, M. Gao, H. Zheng, Y. Chen, X. Wang, H. Bu, F. Song, X. Wan, B. Wang, and G. Wang, Phys. Rev. B **96**, 041201(R) (2017).

[5] M. Koshino and T. Ando, Phys. Rev. B **81**, 195431 (2010).

[6] M. Yan, H. Huang, K. Zhang, E. Wang, W. Yao, K. Deng, G. Wan, H. Zhang, M. Arita, H. Yang, Z. Sun, H. Yao, Y. Wu, S. Fan, W. Duan, and S. Zhou, Nat. Commun. **8**, 257 (2017).

[7] A. Politano, G. Chiarello, B. Ghosh, K. Sadhukhan, C. N. Kuo, C. S. Lue, V. Pellegrini, and A. Agarwal, Phys. Rev. Lett. **121**, 086804 (2018).

[8] T. R. Chang, S. Y. Xu, D. S. Sanchez, W. F. Tsai, S. M. Huang, G. Chang, C. H. Hsu, G. Bian, I. Belopolski, Z. M. Yu, S. A. Yang, T. Neupert, H. T. Jeng, H. Lin, and M. Z. Hasan, Phys. Rev. Lett. **119**, 026404 (2017).

[9] K. Zhang, M. Yan, H. Zhang, H. Huang, M. Arita, Z. Sun, W. Duan, Y. Wu, and S. Zhou, Phys. Rev. B **96**, 125102 (2017).

[10] H. J. Noh, J. Jeong, E. J. Cho, K. Kim, B. I. Min, and B. G. Park, Phys. Rev. Lett. **119**, 016401 (2017).

[11] K. Deng, M. Yan, C.-P. Yu, J. Li, X. Zhou, K. Zhang, Y. Zhao, K. Miyamoto, T. Okuda, W. Duan, Y. Wu, X. Zhong, and S. Zhou, Science Bulletin **64**, 1044 (2019).

[12] M. K. Lin, R. A. B. Villaos, J. A. Hlevyack, P. Chen, R. Y. Liu, C. H. Hsu, J. Avila, S. K. Mo, F. C. Chuang, and T. C. Chiang, Phys. Rev. Lett. **124**, 036402 (2020).

[13] H. Huang, S. Zhou, and W. Duan, Phys. Rev. B **94**, 121117(R) (2016).

[14] W. Zheng, R. Schönemann, N. Aryal, Q. Zhou, D. Rhodes, Y. C. Chiu, K. W. Chen, E. Kampert, T. Förster, T. J. Martin, G. T. McCandless, J. Y. Chan, E. Manousakis, and L. Balicas, Phys. Rev. B **97**, 235154 (2018).
[15] H. Ma, P. Chen, B. Li, J. Li, R. Ai, Z. Zhang, G. Sun, K. Yao, Z. Lin, B. Zhao, R. Wu, X. Tang, X. Duan, and X. Duan, Nano Lett. **18**, 3523 (2018).
[16] S. Hao, J. Zeng, T. Xu, X. Cong, C. Wang, C. Wu, Y. Wang, X. Liu, T. Cao, G. Su, L. Jia, Z. Wu, Q. Lin, L. Zhang, S. Yan, M. Guo, Z. Wang, P. Tan, L. Sun, Z. Ni, S.-J. Liang, X. Cui, and F. Miao, Adv. Funct. Mater. **28**, 1803746 (2018).
[17] Q. Wang, C. Z. Li, S. Ge, J. G. Li, W. Lu, J. Lai, X. Liu, J. Ma, D. P. Yu, Z. M. Liao, and D. Sun, Nano Lett. **17**, 834 (2017).
[18] X. W. Tong, Y. N. Lin, R. Huang, Z. X. Zhang, C. Fu, D. Wu, L. B. Luo, Z. J. Li, F. X. Liang, and W. Zhang, ACS Appl. Mater. Interfaces **12**, 53921 (2020).
[19] T. Liang, Q. Gibson, M. N. Ali, M. Liu, R. J. Cava, and N. P. Ong, Nat. Mater. **14**, 280 (2014).
[20] Y. Wang, E. Liu, A. Gao, T. Cao, M. Long, C. Pan, L. Zhang, J. Zeng, C. Wang, W. Hu, S. J. Liang, and F. Miao, ACS Nano **12**, 9513 (2018).
[21] G. Jnawali, Y. Rao, H. Yan, and T. F. Heinz, Nano Lett. **13**, 524 (2013).
[22] K. J. Tielrooij, J. C. W. Song, S. A. Jensen, A. Centeno, A. Pesquera, A. Zurutuza Elorza, M. Bonn, L. S. Levitov, and F. H. L. Koppens, Nat. Phys. **9**, 248 (2013).
[23] C. H. Lui, A. J. Frenzel, D. V. Pilon, Y. H. Lee, X. Ling, G. M. Akselrod, J. Kong, and N. Gedik, Phys. Rev. Lett. **113**, 166801 (2014).
[24] R. Valdés Aguilar, J. Qi, M. Brahlek, N. Bansal, A. Azad, J. Bowlan, S. Oh, A. J. Taylor, R. P. Prasankumar, and D. A. Yarotski, Appl. Phys. Lett. **106**, 011901 (2015).
[25] K.-C. Lin, M.-Y. Li, D. C. Ling, C. C. Chi, and J.-C. Chen, Phys. Rev. B **91**, 125440 (2015).
[26] G. Jnawali, Y. Rao, H. Yan, and T. F. Heinz, Nano Lett. **13**, 524 (2013).
[27] S. Kar, D. R. Mohapatra, E. Freysz, and A. K. Sood, Phys. Rev. B **90**, 165420 (2014).
[28] C. J. Docherty, C. T. Lin, H. J. Joyce, R. J. Nicholas, L. M. Herz, L. J. Li, and M. B. Johnston, Nat. Commun. **3**, 1228 (2012).
[29] W. Zhang, Y. Yang, P. Suo, W. Zhao, J. Guo, Q. Lu, X. Lin, Z. Jin, L. Wang, G. Chen, F. Xiu, W. Liu, C. Zhang, and G. Ma, Appl. Phys. Lett. **114**, 221102 (2019).
[30] W. Lu, J. Ling, F. Xiu, and D. Sun, Phys. Rev. B **98**, 104310 (2018).
[31] L. Fu, D. Hu, R. G. Mendes, M. H. Rummeli, Q. Dai, B. Wu, L. Fu, and Y. Liu, ACS Nano **12**, 9405 (2018).
[32] O. Pavlosiuk and D. Kaczorowski, Sci. Rep. **8**, 11297 (2018).
[33] J. Lai, J. Ma, Y. Liu, K. Zhang, X. Zhuo, J. Chen, S. Zhou, and D. Sun, 2D Materials **7**, 034003 (2020).
[34] M. S. Shawkat, T. A. Chowdhury, H. S. Chung, S. Sattar, T. J. Ko, J. A. Larsson, and Y. Jung, Nanoscale **12**, 23116 (2020).
[35] H. Xu, C. Guo, J. Zhang, W. Guo, C. N. Kuo, C. S. Lue, W. Hu, L. Wang, G. Chen, A. Politano, X. Chen, and W. Lu, Small **15**, e1903362 (2019).
[36] L. Zeng, D. Wu, J. Jie, X. Ren, X. Hu, S. P. Lau, Y. Chai, and Y. H. Tsang, Adv. Mater, e2004412 (2020).
[37] J. Lu, H. Liu, and J. Sun, Nanotechnology **28**, 464001 (2017).
[38] R. H. M. Groeneveld, R. Sprik, and A. Lagendijk, Phys. Rev. B Condens. Matter. **51**, 11433 (1995).

[39] Y. Ishida, H. Masuda, H. Sakai, S. Ishiwata, and S. Shin, Phys. Rev. B **93**, 100302(R) (2016).
[40] E. Baek, T. Rim, J. Schutt, C. K. Baek, K. Kim, L. Baraban, and G. Cuniberti, Nano Lett. **17**, 6727 (2017).
[41] Y. Yang, X. Peng, H. S. Kim, T. Kim, S. Jeon, H. K. Kang, W. Choi, J. Song, Y. J. Doh, and D. Yu, Nano Lett. **15**, 5875 (2015).
[42] D. Khokhlov, L. Ryabova, A. Nicorici, V. Shklover, S. Ganichev, S. Danilov, and V. Bel'kov, Appl. Phys. Lett. **93**, 264103 (2008).
[43] B. A. Akimov and V. A. Bogoyavlenskiy, Phys. Rev. B **61**, 16045 (2000).
[44] R. Sreekumar, R. Jayakrishnan, C. Sudha Kartha, and K. P. Vijayakumar, J. Appl. Phys. **100**, 033707 (2006).
[45] M. Ziwritsch, S. Müller, H. Hempel, T. Unold, F. F. Abdi, R. van de Krol, D. Friedrich, and R. Eichberger, ACS Energy Lett. **1**, 888 (2016).
[46] A. J. Rettie, W. D. Chemelewski, D. Emin, and C. B. Mullins, J. Phys. Chem. Lett. **7**, 471 (2016).
[47] D. Venkateshvaran, M. Nikolka, A. Sadhanala, V. Lemaur, M. Zelazny, M. Kepa, M. Hurhangee, A. J. Kronemeijer, V. Pecunia, I. Nasrallah, I. Romanov, K. Broch, I. McCulloch, D. Emin, Y. Olivier, J. Cornil, D. Beljonne, and H. Sirringhaus, Nature **515**, 384 (2014).
[48] E. Cinquanta, D. Meggiolaro, S. G. Motti, M. Gandini, M. J. P. Alcocer, Q. A. Akkerman, C. Vozzi, L. Manna, F. De Angelis, A. Petrozza, and S. Stagira, Phys. Rev. Lett. **122**, 166601 (2019).
[49] S. A. Bretschneider, I. Ivanov, H. I. Wang, K. Miyata, X. Zhu, and M. Bonn, Adv. Mater, e1707312 (2018).
[50] D. M. Kiyoshi Miyata, M. Tuan Trinh, Prakriti P. Joshi, Edoardo Mosconi, and F. D. A. Skyler C. Jones, X.-Y. Zhu, Sci. Adv, **3**, e1701217 (2017).
[51] C. Kenfack-Sadem, M. F. C. Fobasso, F. Amo-Mensah, E. Baloitcha, A. Fotué, and L. C. Fai, Physica E: Low-dimensional Systems and Nanostructures **122**, 114154 (2020).
[52] B. S. Kandemir, J. Phys. Condens. Matter. **25**, 025302 (2013).
[53] Q. Chen, W. Wang, and F. M. Peeters, J. Appl. Phys. **123**, 214303 (2018).
[54] G. Anemone, M. Garnica, M. Zappia, P. C. Aguilar, A. Al Taleb, C.-N. Kuo, C. S. Lue, A. Politano, G. Benedek, A. L. V. de Parga, R. Miranda, and D. Farías, 2D Mater. **7**, 025007 (2020).
[55] P. B. Miranda, D. Moses, and A. J. Heeger, Phys. Rev. B **64**, 4758 (2001).

<u>**Supplemental Material**</u>

## Note 1. The thickness characterization of $PtTe_2$ films on fused silica substrate by atomic force microscopy (AFM).

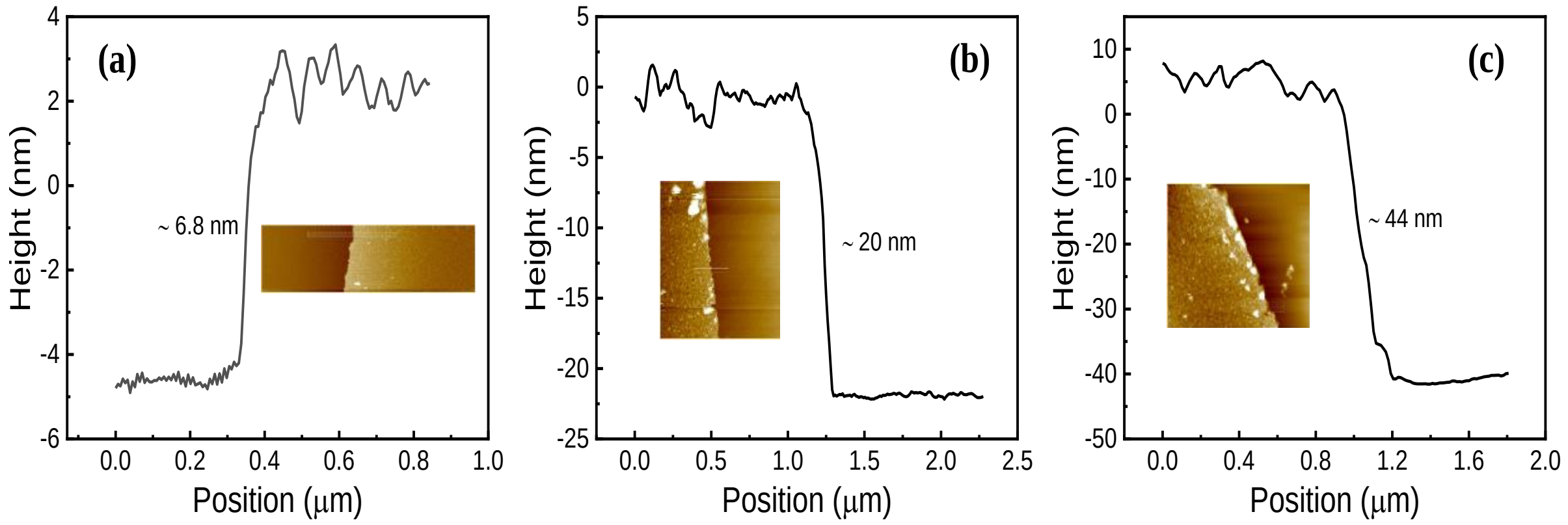

Figure S1. From the AFM topography maps it is seen that the thickness of $PtTe_2$ film is around (a) 6.8 nm, (b) 20 nm and (c) 44 nm, respectively.

## Note 2. XPS spectra of three-thickness $PtTe_2$ films on fused silica substrate.

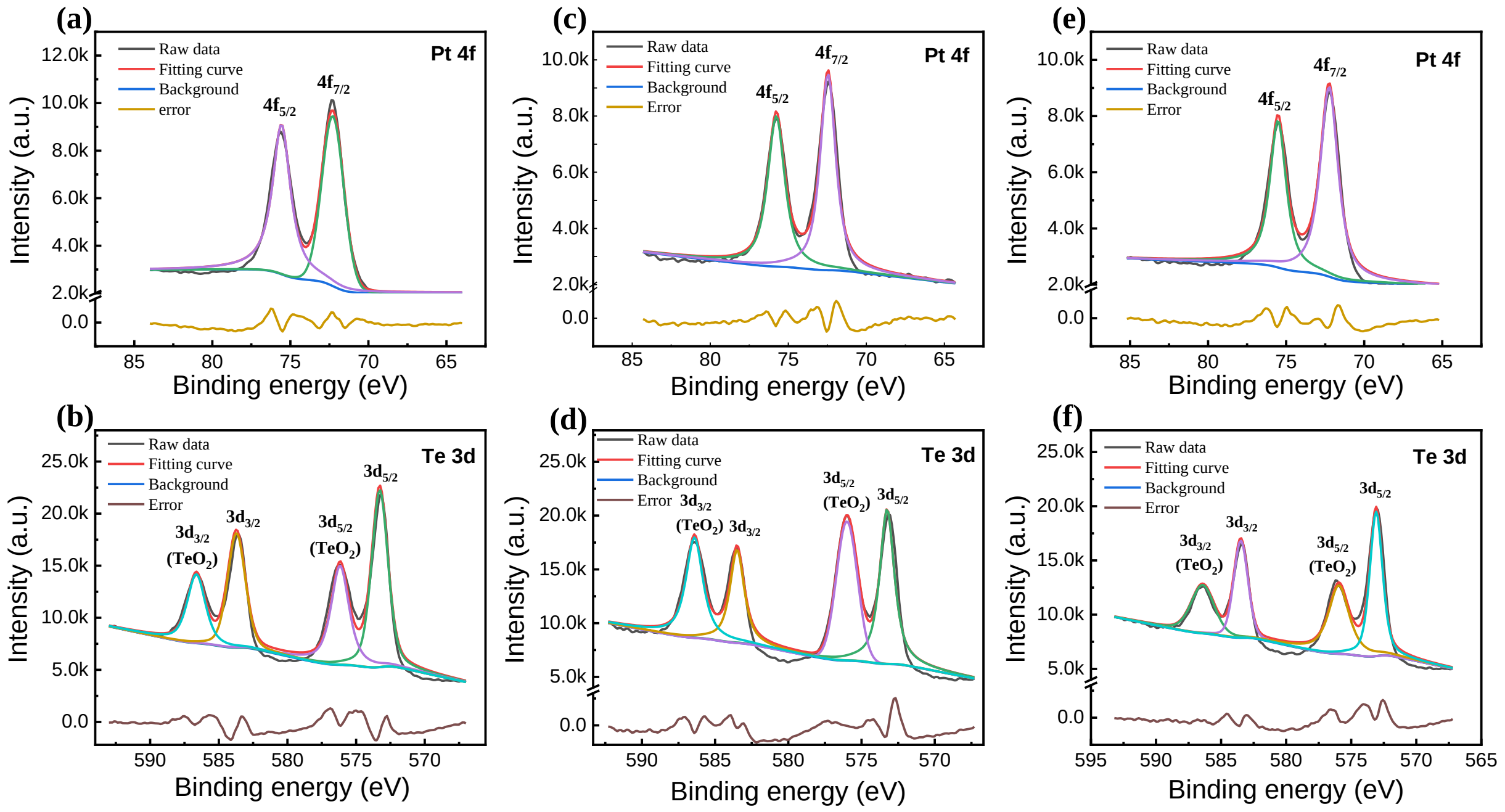

Figure S2. High-resolution Pt 4f and Te 3d XPS spectra of $PtTe_2$ films with different thickness of (a) (b) 6.8 nm, (c) (d) 20 nm and (e) (f) 44 nm. For Pt 4f core levels, two distinct peaks at ≈72.4 and ≈75.8 eV attributed to the Pt $4f_{7/2}$ and Pt $4f_{5/2}$ states, respectively. Similarly, two primary peaks at

≈573.1 and ≈583.5 eV represent the Te $3d_{5/2}$ and Te $3d_{3/2}$ states, and other two fragile ones at ≈576.2 and ≈586.5 eV can be assigned to $TeO_2$.

**Note 3. Carrier concentrations for 6.8, 20 and 44 nm $PtTe_2$ films at room temperature are measured with fabricated Hall bar devices. The calculated carrier concentrations of $PtTe_2$ films are around 3.3 × $10^{22}$ $cm^{-3}$, 1.4 × $10^{22}$ $cm^{-3}$ and 1.2 × $10^{22}$ $cm^{-3}$ for 6.8, 20 and 44 nm films, respectively.**

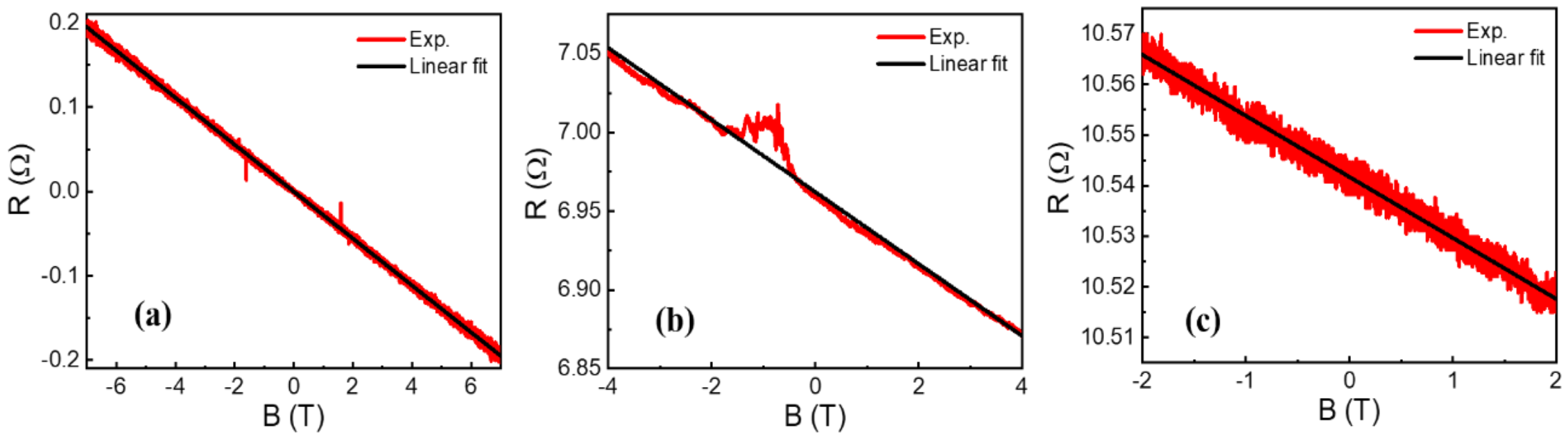

Figure S3. The Hall R-H curves of $PtTe_2$ films with thickness of (a) 6.8 nm, (b) 20 nm and (c) 44 nm; the red line denotes experimental data, and the black line denotes linear fitting curve.

**Note 4. Transient THz dynamics of 6.8 nm and 20 nm $PtTe_2$ films on sapphire substrate at room temperature.**

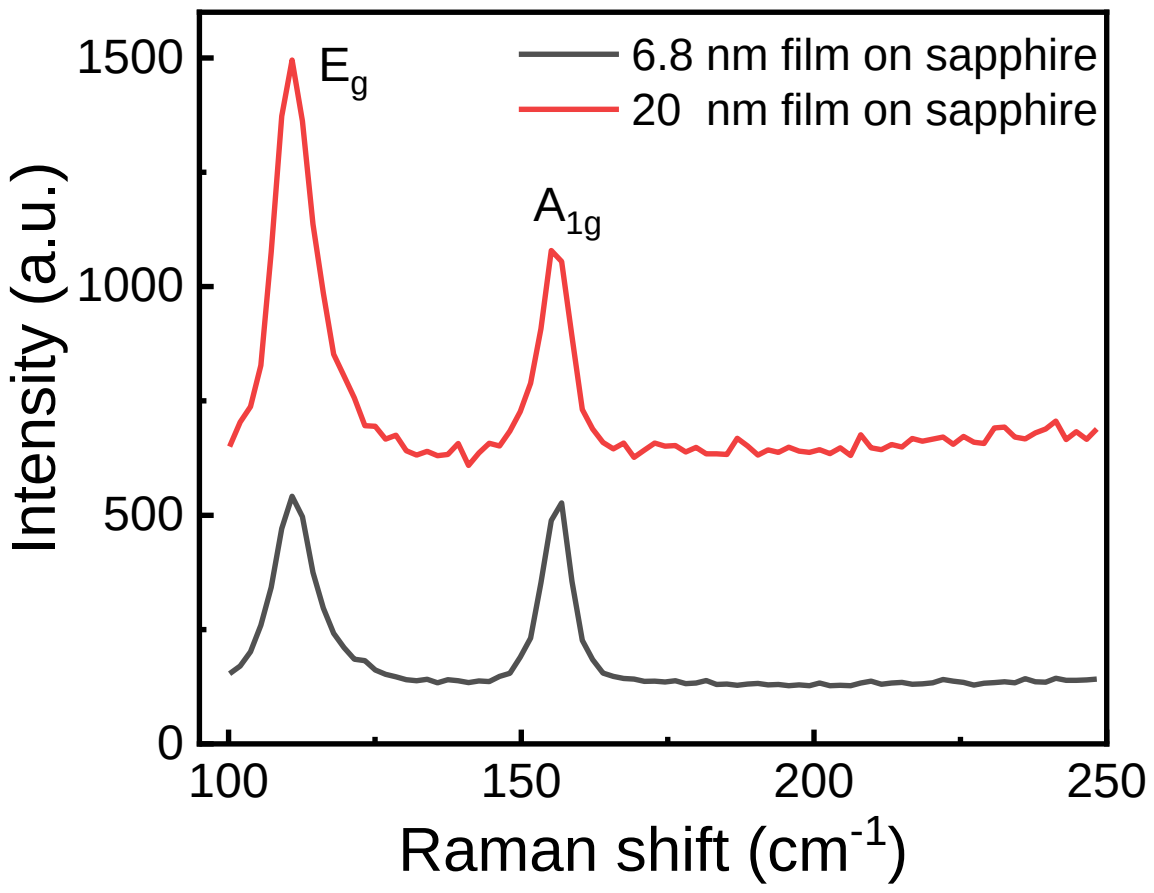

Figure S4.1. Raman spectra of 6.8 nm and 20 nm $PtTe_2$ films on sapphire substrate.

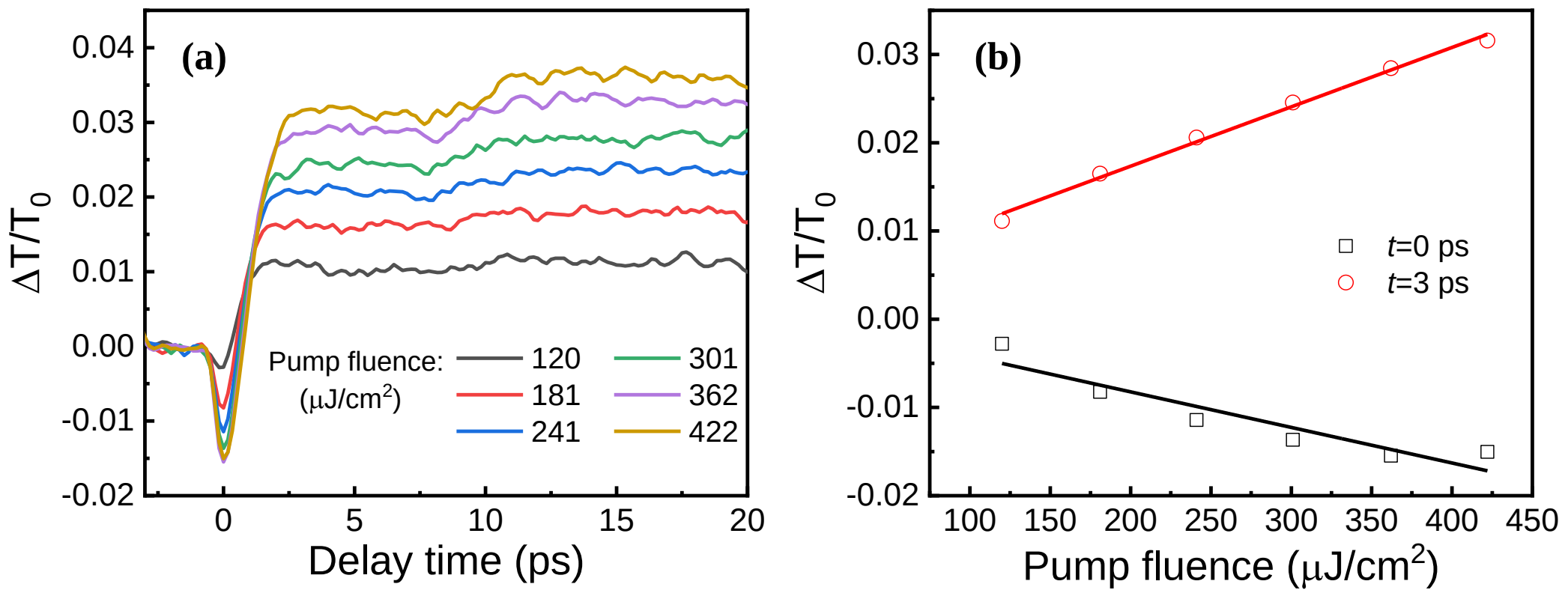

Figure S4.2. (a) Transient THz transmission of 6.8 nm $PtTe_2$ film on sapphire substrate. (b) The magnitude of $\Delta T/T_0$ at $t$=0 ps (black) and $t$=3 ps (red) with respect to the pump fluence, respectively.

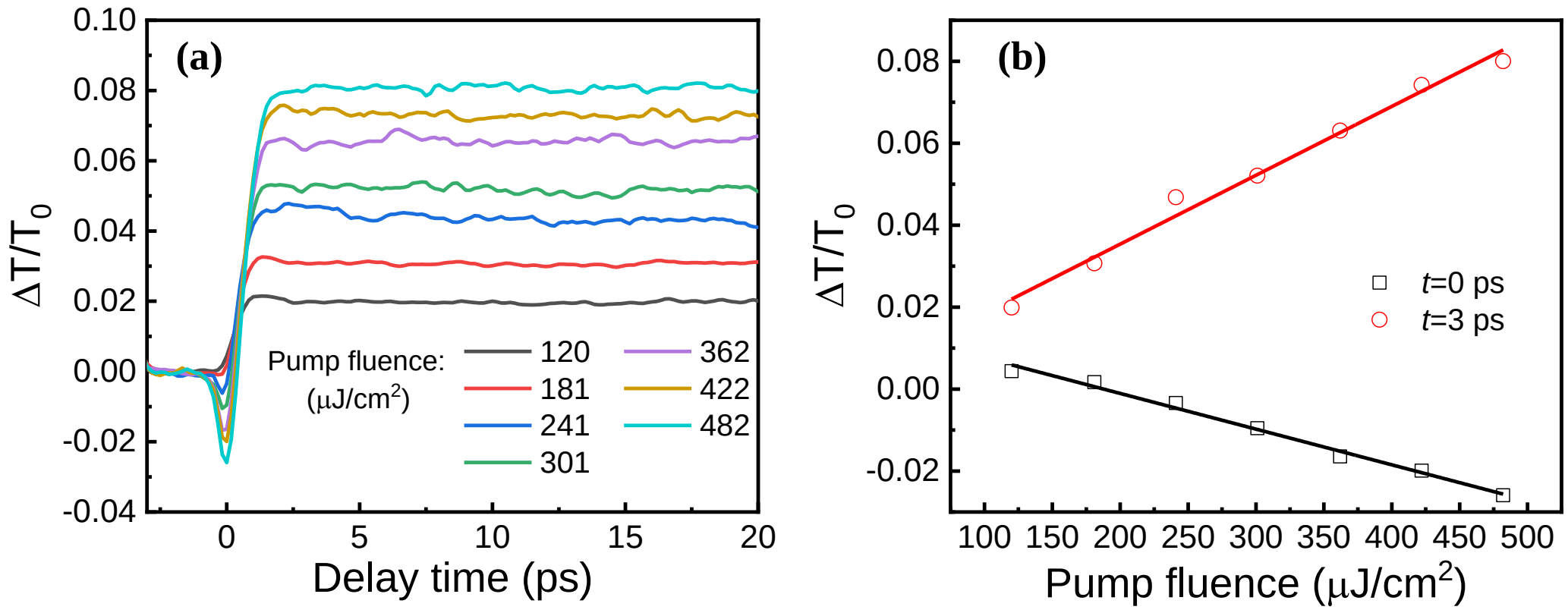

Figure S4.3. (a) Transient THz transmission of 20 nm $PtTe_2$ film on sapphire substrate. (b) The magnitude of $\Delta T/T_0$ collected at $t$=0 ps (black) and $t$=3 ps (red) with respect to the pump fluence, respectively.

## Note 5. Transient THz dynamics of 6.8 nm and 20 nm $PtTe_2$ films on YAG substrate at room temperature.

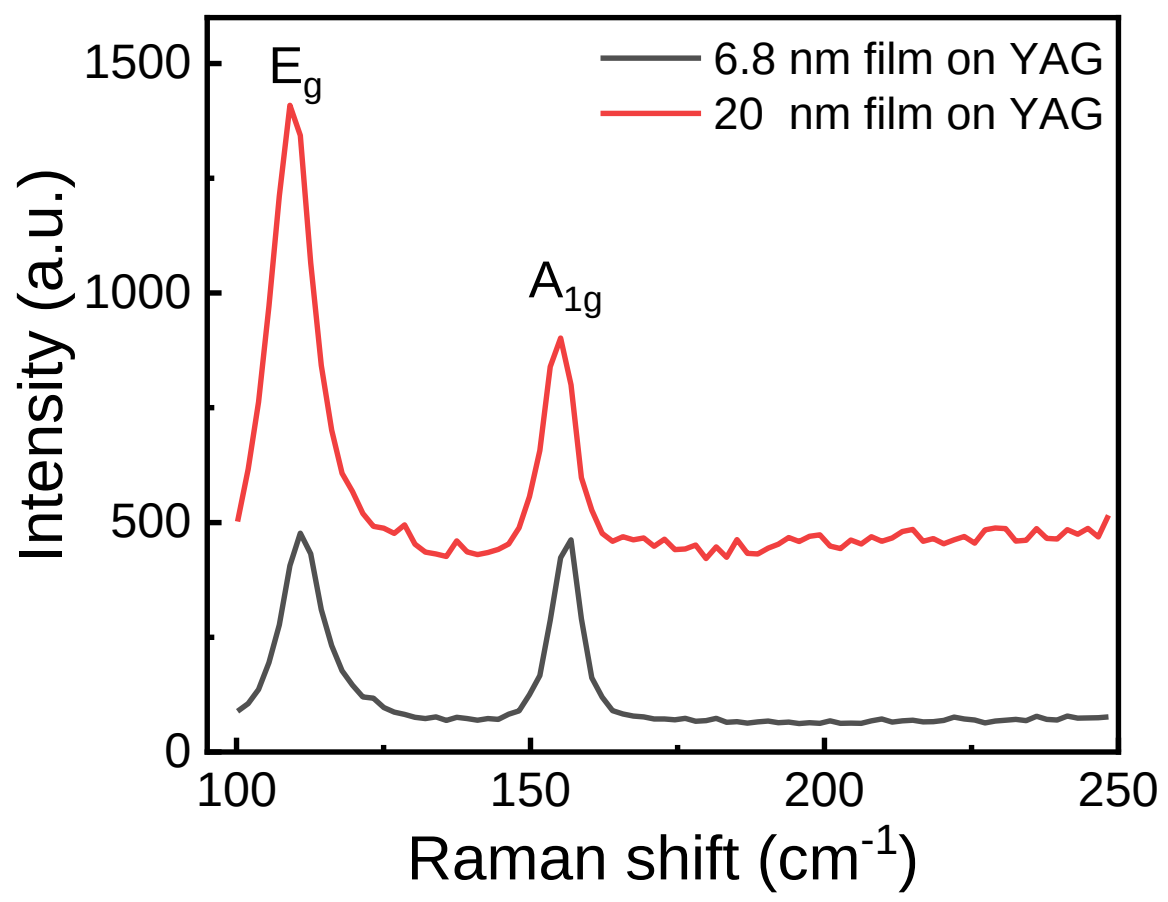

Figure S5.1. Raman spectra of 6.8 nm and 20 nm $PtTe_2$ films on YAG substrate.

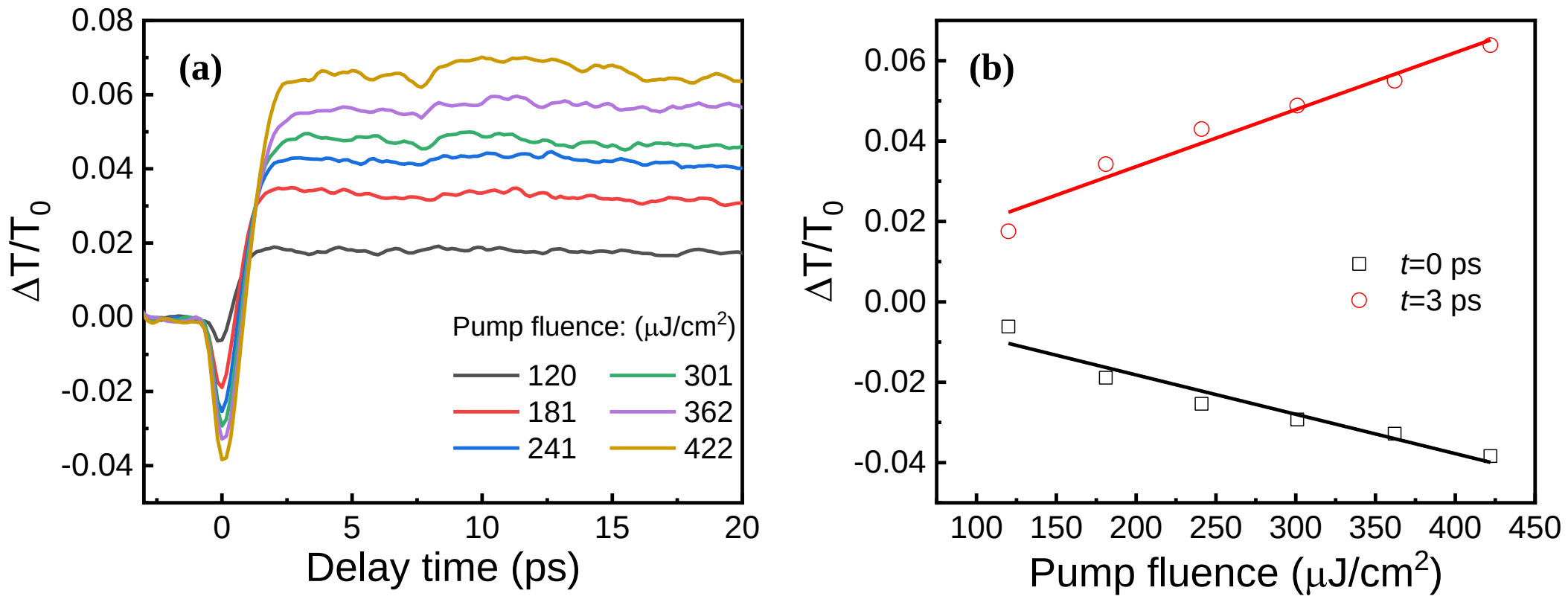

Figure S5.2. (a) Transient THz transmission of 6.8 nm $PtTe_2$ film on YAG substrate. (b) The magnitude of $\Delta T/T_0$ collected at $t$=0 ps (black) and $t$=3 ps (red) with respect to the pump fluence, respectively.

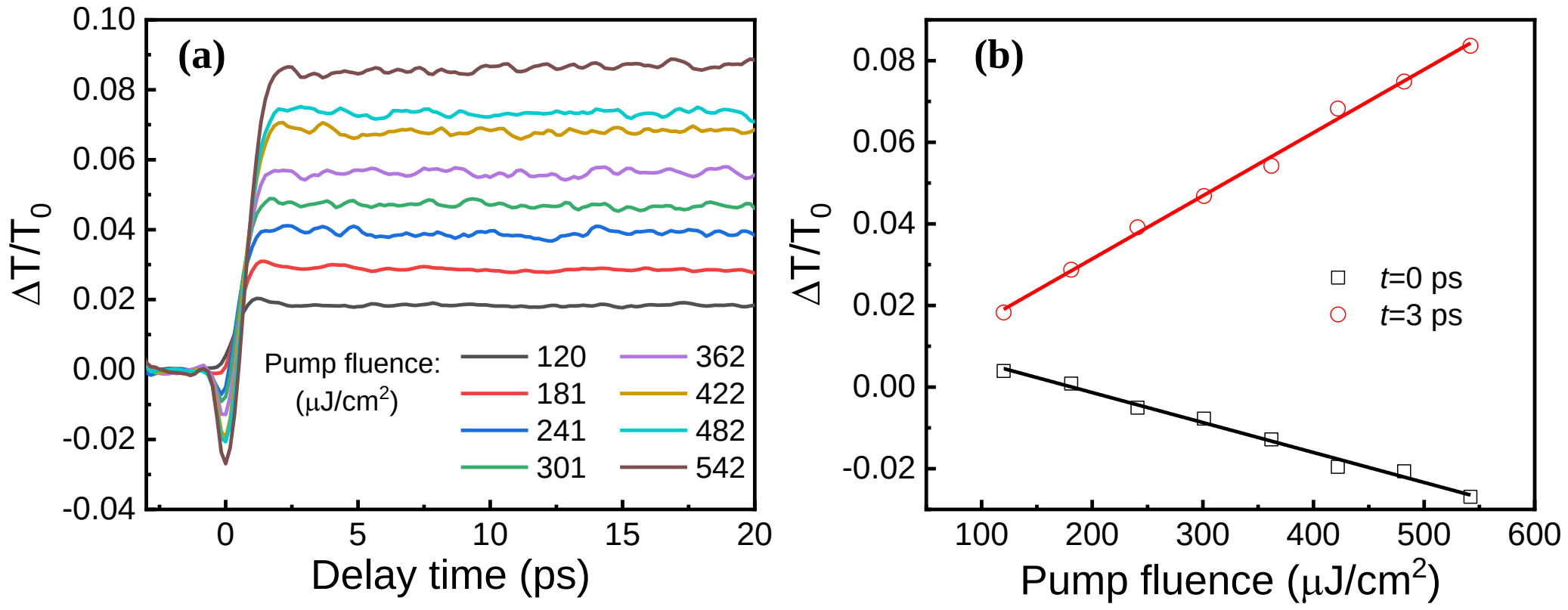

Figure S5.3. (a) Transient THz transmission of 20 nm $PtTe_2$ film on YAG substrate. (b) The magnitude of $\Delta T/T_0$ collected at $t$=0 ps (black) and $t$=3 ps (red) with respect to the pump fluence,

respectively.

**Note 6. To quantitatively understand the transient dynamics of 20 nm $PtTe_2$ film, the convoluted monoexponential model with the following form [1,2]was employed to fit the ultrafast transient THz transmission response in the first 10 ps time window using Origin 2018 software package.**

$$\frac{\Delta T}{T_0}(t) = A * \exp\left[\left(\frac{\omega}{\tau}\right)^2 - \frac{t}{\tau}\right] * \left[1 - \mathrm{Erf}\left(\frac{\omega}{\tau} - \frac{t}{2\omega}\right)\right] + B$$

Where $t$ is the pump-probe delay time, $\tau$ and A are the correlative relaxation constant and amplitude, respectively. ω is the THz probe pulse, and B is the time-independent offset.

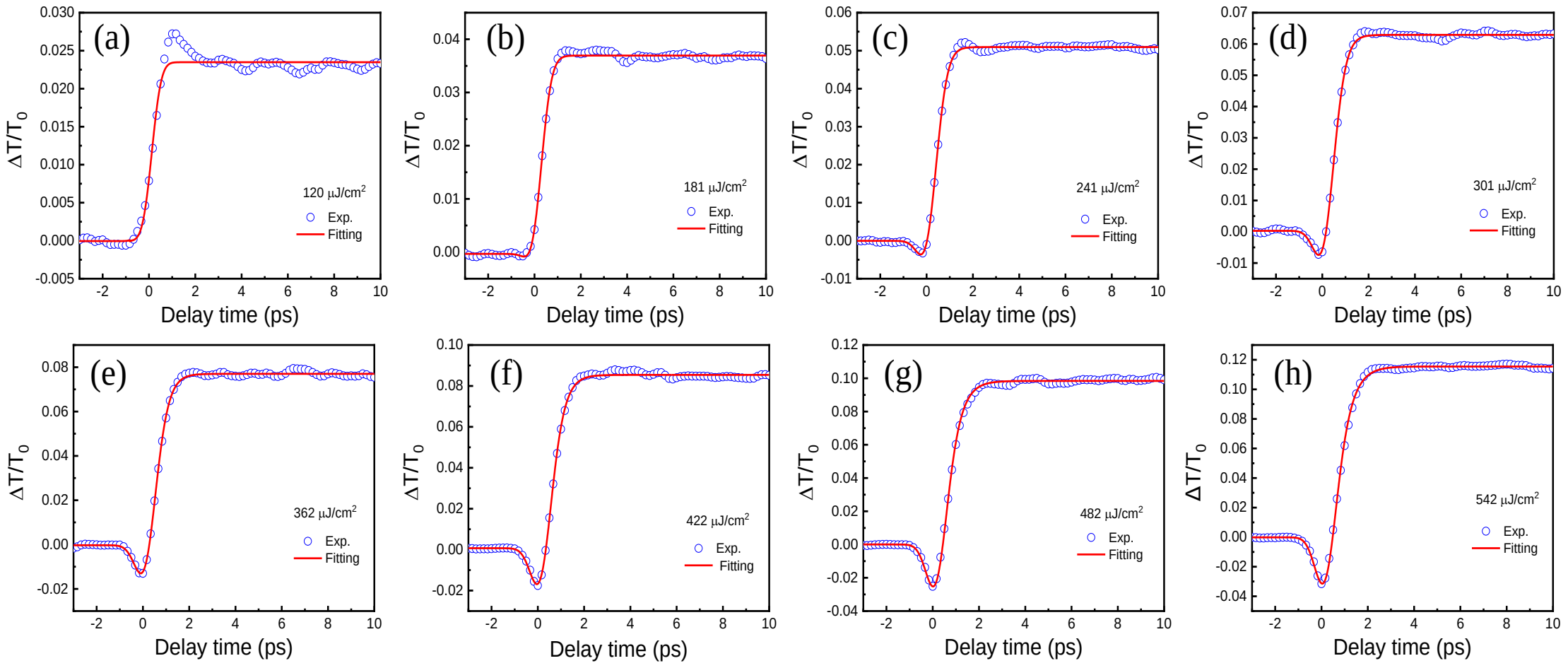

Figure S6. The transient dynamics under 780 nm optical pump at room temperature and the corresponding fitted curves with convoluted monoexponential decay function under various pump fluences; the blue dots denote experimental data, and the red lines denote fitted curves.

**Note 7. Discussions on possible contributions to the observed negative THz photoconductivity.**

Here we assigned the observed negative THz PC as the excitation of polaron in 20 nm $PtTe_2$ film. In order to exclude other possible contributions on the NPC, we listed and discussed each mechanism as follows:

**Interband scattering.** The electron in the conduction band has much larger effective mass than that in Dirac cone on account of the fact that Dirac fermion is described in massless Dirac equation [3,4], which gives rise to the electron mobility in conduction band is much smaller than that in Dirac cone. Therefore, the photobleaching behavior may originate from the sharp decline of electron mobility as the electrons in Dirac cones are photoexcited into conduction band with a 1.59-eV-optical-pump. However, the contribution of the electrons in the Dirac cones to conductivity can be ignored due to the fact that the Dirac point of $PtTe_2$ is located around 0.8 eV below the Fermi surface [5,6]. Additionally, we also carried out OPTP measurement with pump wavelength of 1300 nm and 1600 nm, the pump-induced bleaching of THz transmission is still clearly observed as shown in Figure S7.1. Therefore, the conjecture about the contribution of Dirac electrons can be easily excluded.

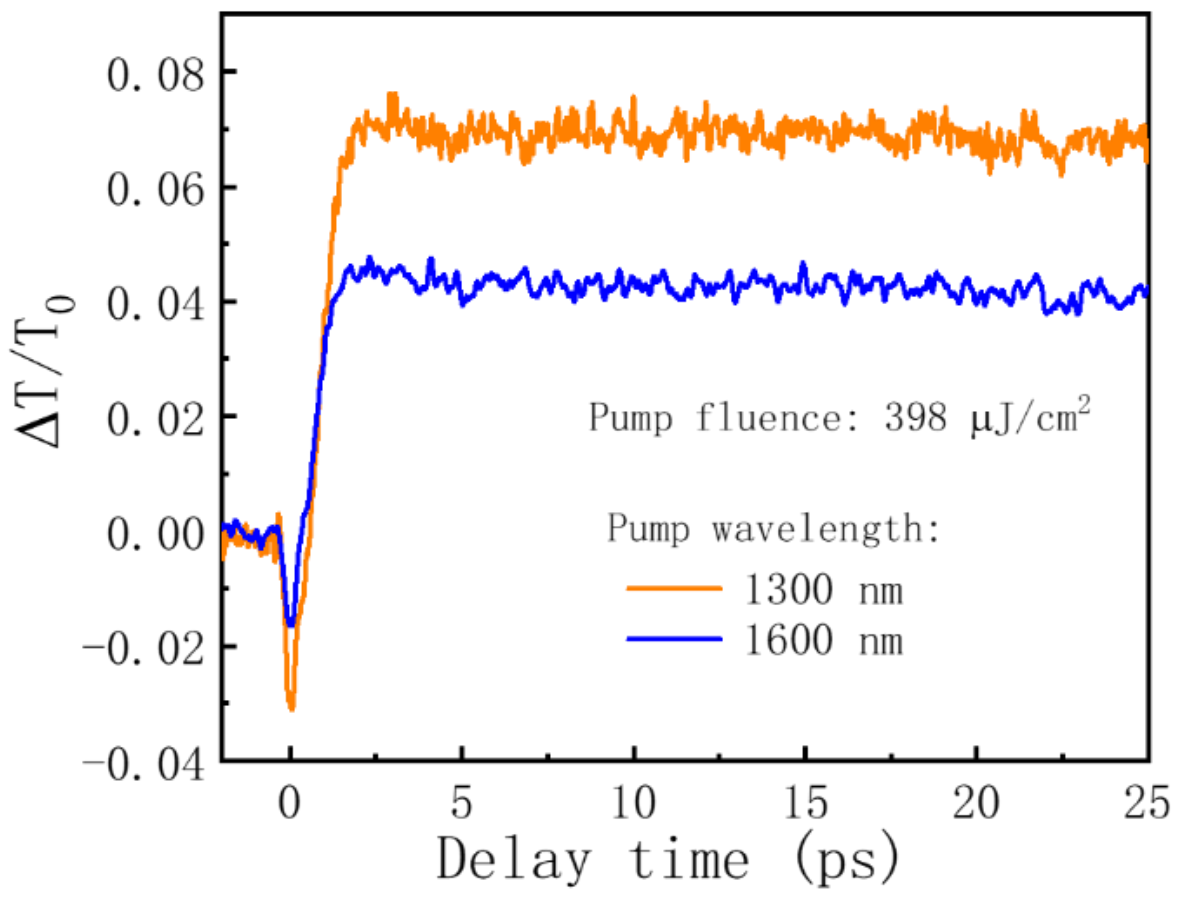

**Figure S7.1.** The transient THz transmission change under 1300 nm (0.95 eV) and 1600 nm (0.775 eV) optical pump under pump fluence of 377 $\mu J/cm^2$ at room temperature.

**The formation of trions.** Another possibility that leads to negative THz PC is the formation of trions, i.e. the charged three-particle state, which come from that the excessive free electrons are captured by photogenerated electron-hole pairs [7,8]. The charged quasiparticles, i.e. the trions, with increased effective mass have lower carrier mobility, hence causing the diminution of the film's conductivity. Consider the metal-like nature of 20 nm $PtTe_2$ film with much high carrier concentration, the formation of the exciton with photoexcitation can be totally screened by the background free carriers, we can safely rule out this possibility.

**The two-temperature model (TTM).** The TTM has been proved to be acceptable model in some metals or semimetals. The TTM assumes that after photoexcitation the e-e thermalization (in tens of fs) is much faster than e-ph thermalization with a typical time of ~1.0 ps, and the electron with elevated temperature interacts with cold lattice via e-ph coupling to transfer the excess energy to lattice until the two subsystems reach a balanced temperature [9-12]. It seems to be reasonable in interpreting the occurrence of bleaching signal because the observed subpicosecond relaxation time from positive to negative of THz PC is consistent with the character time of e-ph coupling in TTM. However, e-ph coupling leads to the cooling of hot electrons until reach a thermal equilibrium between electron and lattice, and the electronic temperature would not be much lower than that before photoexcitation. Obviously, the TTM fails in account for our experimental phenomenon.

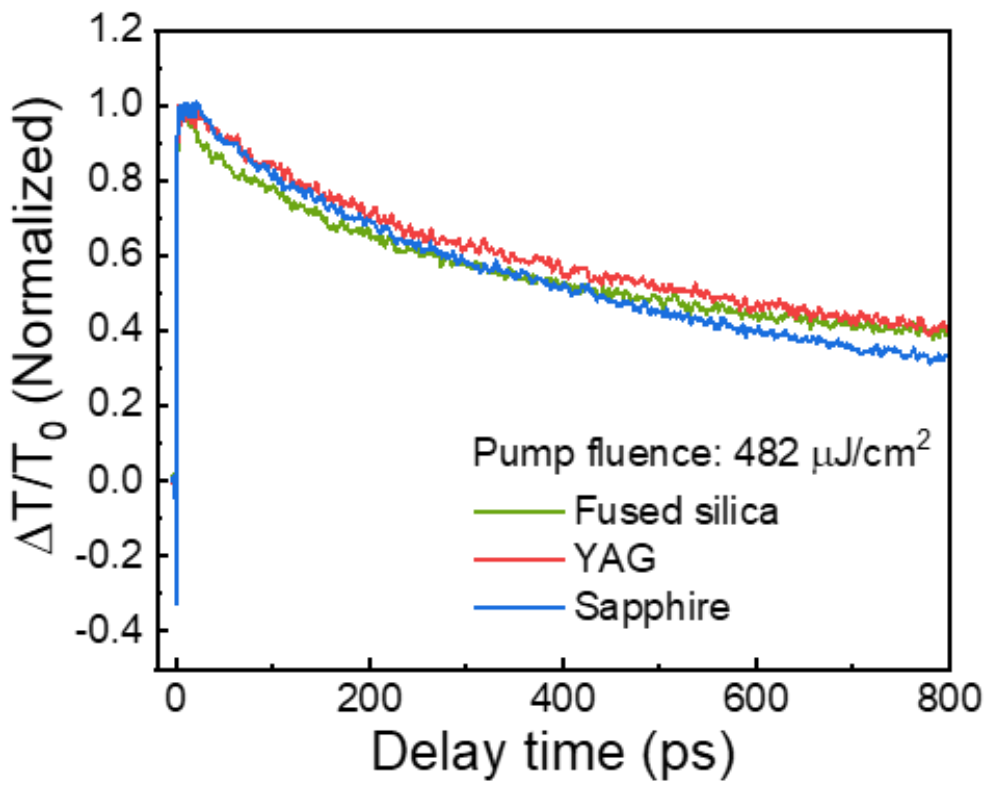

Figure S7.2. Normalized transient THz transmission of 20 nm $PtTe_2$ on fused silica (green), YAG (red) and sapphire (blue) substrates with pump fluence of 482 μJ/cm$^2$ at room temperature.

**Impurity contribution to the NPC:** Impurity in the $PtTe_2$ film could also lead to the NPC after photoexcitation as reported in some semiconductors [13-15]. For the case of across bandgap photoexcitation, photogenerated electrons can be trapped by the defect states (impurity), and the hole left in valance band can recombine with free electrons, as a result, the total carrier population is reduced after photoexcitation, which leads to the NPC. Here we exclude the contribution of impurity to the NPC is based on the following two facts: the first one is the temperature dependent THz conductivity in absence of pump illumination. Figure S7.2 shows the THz transmission change with respect to temperature, it is seen the THz peak-to-peak transmission decreases continuously with decreasing temperature in the investigated temperature, which indicate the conductivity of the films increase with decreasing temperature. The temperature dependent THz conductivity manifests the metallic nature of our $PtTe_2$ film, and the THz conductivity is mainly contributed by free carriers of the films, and the impurity contribution is negligible because the conductivity contributed by impurity conventionally shows increase with temperature. The other is that the relaxation time of the NPC lasts hundreds of ps as shown in Fig. 2(a), which is 4~6 orders of magnitude faster than the recovering time of NPC in semiconductors [13,16,17].

For the 20 nm $PtTe_2$ film, the transmission change of THz peak-to-peak value from 5 K to 300 K is $\frac{16.59585-8.63736}{8.63736} \times 100\% = 92.14\%$, from which it is seen that the 20-nm $PtTe_2$ film show a good metallic-like property. In addition, we have checked temperature dependent THz transmission through the fused silica substrates, which show almost temperature independence.

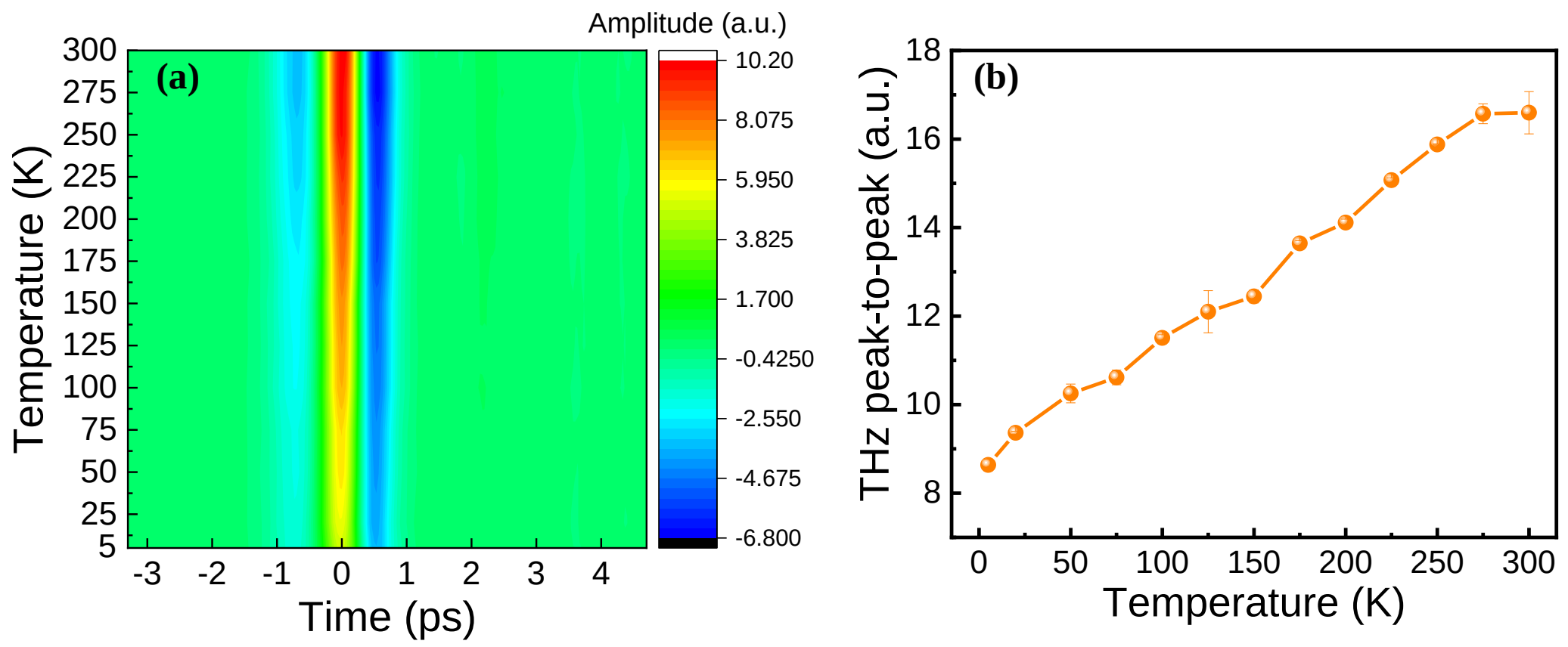

Figure S7.3. The temperature dependent THz transmission of 20 nm $PtTe_2$ film. (a) The transmitted THz mapping of 20 nm film in time domain with various temperatures. (b) The THz peak-to-peak value extracted from (a) with respect to temperature. The orange solid line in (b) give a guide to the eyes.

**Photoinduced THz emission**: The negative THz PC could arise from the photoinduced THz radiation on the surface of $PtTe_2$ film. In order to exclude the contribution from THz radiation, we have carried out THz radiation experiment in transmission configuration with various pump fluences and temperatures under 780 nm optical pump, no clear THz emission signal was observed for the 20 nm $PtTe_2$ film, therefore we can safely rule out the contribution of photo-induced THz radiation to the photobleaching signal.

## Note 8: The transient THz dynamics of 20 nm $PtTe_2$ film on fused silica substrate at various temperature.

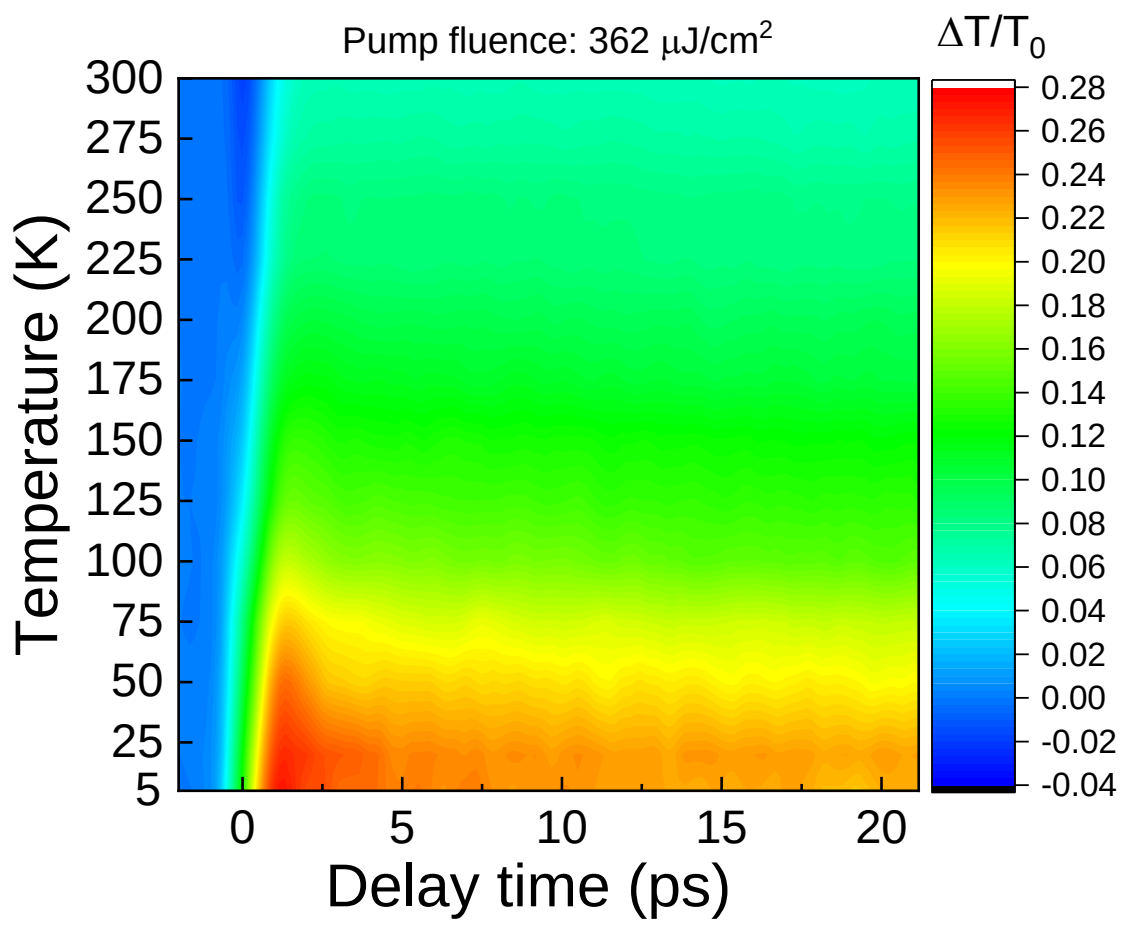

Figure S8.1. The transient THz dynamical spectrum of 20 nm $PtTe_2$ film at various temperature with a fixed pump fluence of 362 μJ/cm$^2$.

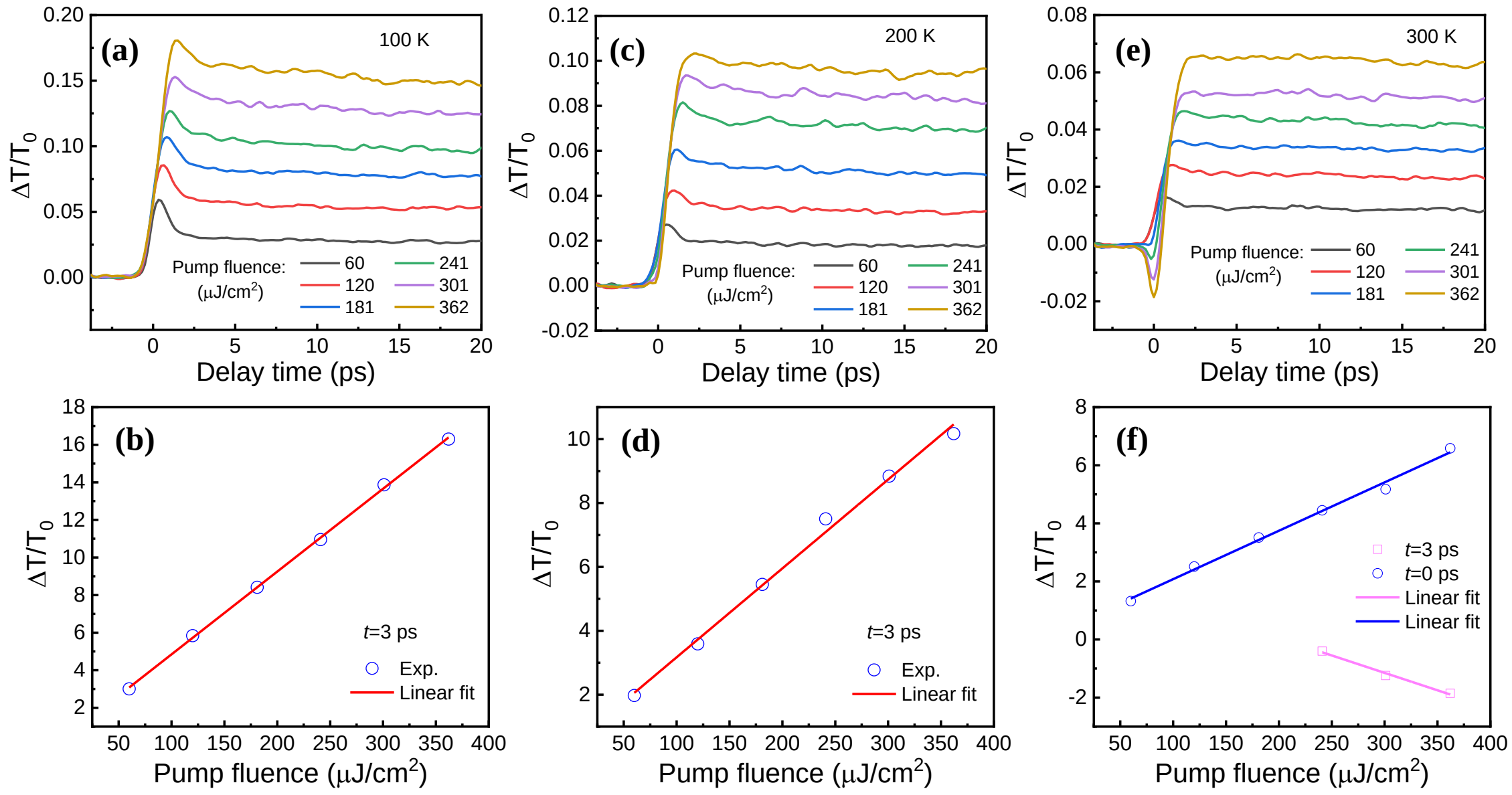

Figure S8.2. Pump fluence dependence of transient THz dynamics of 20 nm $PtTe_2$ film at different temperature. Pump fluence dependent THz transmission change traces at (a) 100 K, (c) 200 K and (e) 300 K, respectively. The magnitude of $\Delta T/T_0$ collected at $t$=0 ps and $t$=3 ps as a function of pump

fluence at (b) 100 K, (d) 200 K and (f) 300 K, respectively.

## Note 9: The fitting details of 6.8 nm, 20 nm and 44 nm film with the same absorbed photon density of $3.2\times10^{14}$ $cm^{-2}$ under 780 nm optical pump at room temperature.

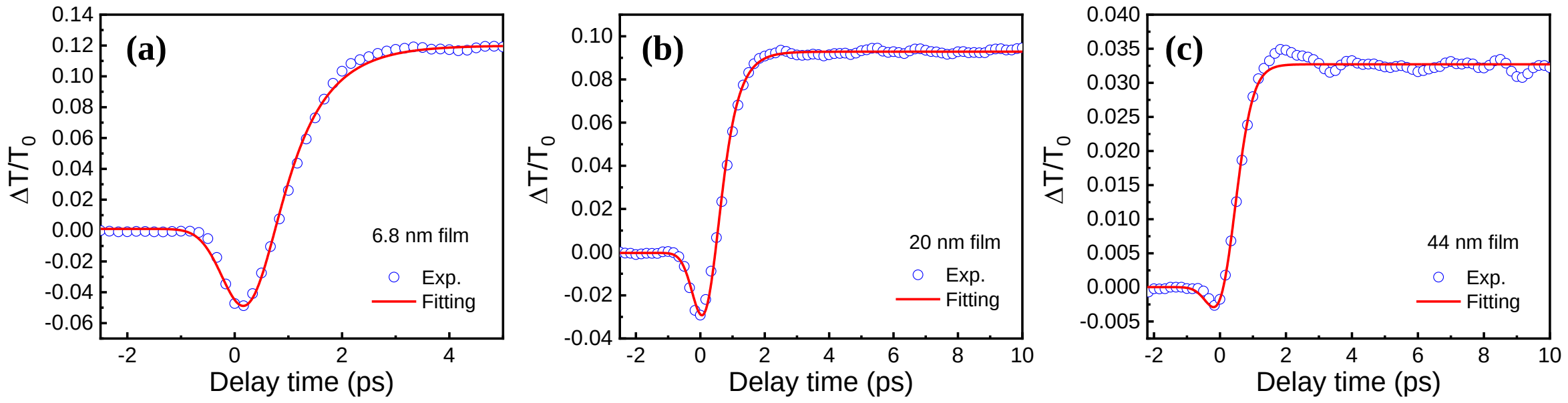

Figure S9. The transient THz dynamics of (a) 6.8 nm, (b) 20 nm and (c) 44 nm $PtTe_2$ films and the corresponding fitted curves with convoluted monoexponential decay function; the blue dots denote experimental data, and the red lines denote fitted curves.

## Note 10: Pump fluence and temperature dependent carrier mobility of $PtTe_2$ film in THz frequency.

Based on THz-TDS measurement without pump, the average conductivity $\sigma_0$ in investigated THz frequency is obtained with thin-film approximation [18,19]

$$\sigma_0 \approx \frac{n_{sub}+1}{Z_0 d}\left(1-\frac{T_0}{T_{sub}}\right)$$

where $n_{sub}$=1.95 is the refractive index for fused silica substrate in THz frequency; $Z_0$=377 Ω is the free space impedance; $d$ =20 nm is the film's thickness; $T_0$ and $T_{sub}$ is the peak value in the THz waveforms through the sample and substrate, respectively. And then the static mobility $\mu_0$ is obtained according to $\sigma_0 = Ne\mu_0$.

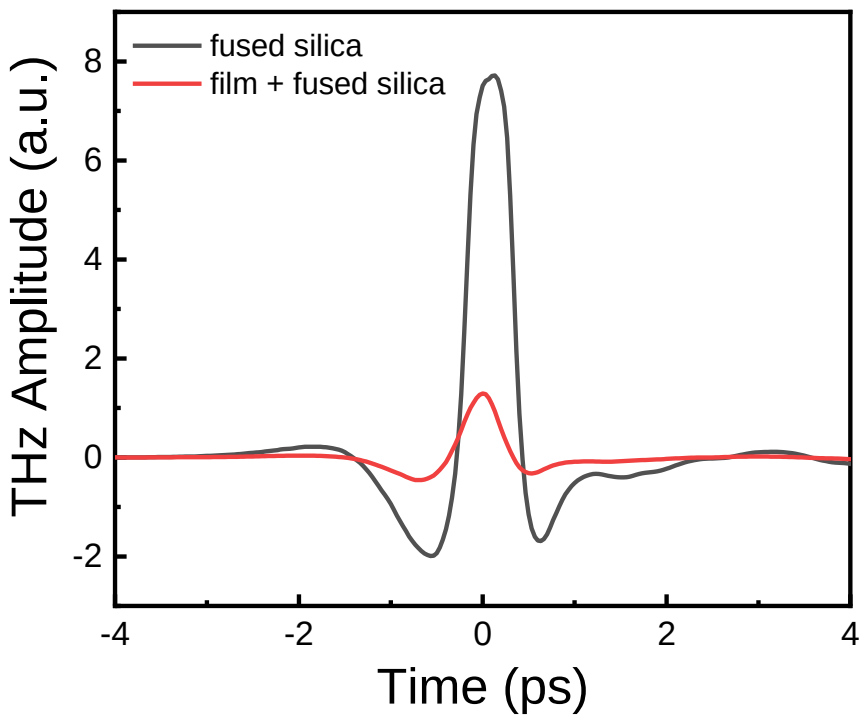

Figure S10.1. The THz-TDS spectra of 20 nm $PtTe_2$ film on fused silica and pure fused silica substrate.

When the pump light is on, pump-induced change in the THz conductivity is given by

$$\Delta\sigma \approx -\frac{n_{sub}+1}{Z_0 d}\frac{\Delta T}{T_0}$$

with ΔT being the change of THz peak value after optical excitation. Due to the negligible increase of photocarriers, we can regard as carrier concentration N almost remain unchanged. So the conductivity change $\Delta\sigma = \sigma - \sigma_0 = \Delta(Ne\mu) = Ne\Delta\mu = Ne(\mu-\mu_0)$, and we can gain the pump fluence and temperature dependent carrier mobility via $\mu = \mu_0 + \Delta\mu$.

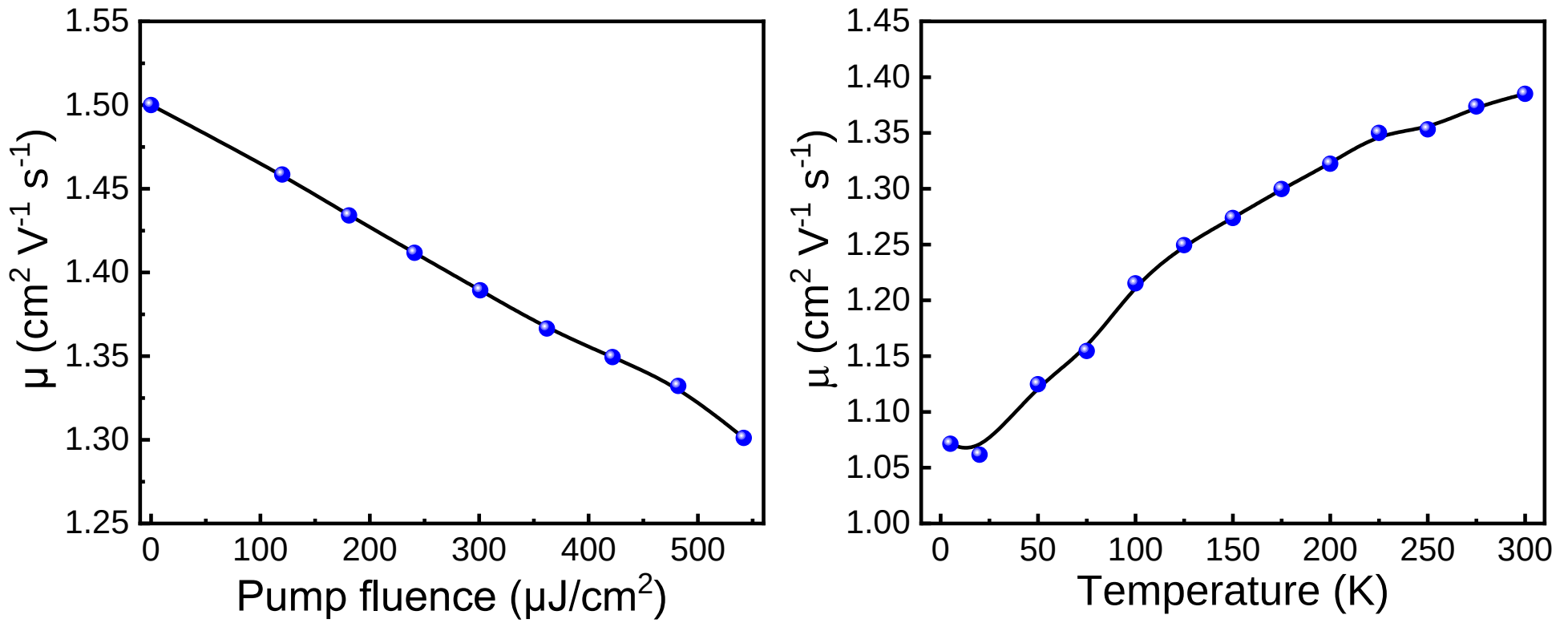

Figure S10.2. (a) The carrier mobility with respect to pump fluence of 20 nm $PtTe_2$ film at pump-probe delay time $t$=3 ps. (b) The temperature dependent carrier mobility of 20 nm $PtTe_2$ film at pump-probe delay time $t$=3 ps under the same pump fluence of 362 μJ/cm$^2$. The blue dots denote experimental data, and the black lines give a guide to the eyes.

## References:

[1] A. V. Kimel, F. Bentivegna, V. N. Gridnev, V. V. Pavlov, R. V. Pisarev, and T. Rasing, Phys. Rev. B **63**, 235201 (2001).

[2] W. Zhang, J. Guo, P. Suo, L. Lv, J. Liu, X. Lin, Z. Jin, W. Liu, and G. Ma, Appl. Opt. **58**, 8200 (2019).

[3] J. C. Smith, S. Banerjee, V. Pardo, and W. E. Pickett, Phys. Rev. Lett. **106**, 056401 (2011).

[4] N. P. Armitage, E. J. Mele, and A. Vishwanath, Rev. Mod. Phys. **90**, 015001 (2018).

[5] M. Yan, H. Huang, K. Zhang, E. Wang, W. Yao, K. Deng, G. Wan, H. Zhang, M. Arita, H. Yang, Z. Sun, H. Yao, Y. Wu, S. Fan, W. Duan, and S. Zhou, Nat. Commun. **8**, 257 (2017).

[6] A. Politano, G. Chiarello, B. Ghosh, K. Sadhukhan, C. N. Kuo, C. S. Lue, V. Pellegrini, and A. Agarwal, Phys. Rev. Lett. **121**, 086804 (2018).

[7] F. P. Daniele Sanvitto, Andrew J. Shields, Peter C. M. Christianen, Stuart N. Holmes, and D. A. R. Michelle Y. Simmons, Jan C. Maan, Michael Pepper, Science **294**, 837 (2001).
[8] C. H. Lui, A. J. Frenzel, D. V. Pilon, Y. H. Lee, X. Ling, G. M. Akselrod, J. Kong, and N. Gedik, Phys. Rev. Lett. **113**, 166801 (2014).
[9] R. H. Groeneveld, R. Sprik, and A. Lagendijk, Phys. Rev. B. Condens Matter **51**, 11433 (1995).
[10] M. Hase, K. Ishioka, J. Demsar, K. Ushida, and M. Kitajima, Phys. Rev. B **71**, 184301 (2005).
[11] Y. M. Dai, J. Bowlan, H. Li, H. Miao, S. F. Wu, W. D. Kong, Y. G. Shi, S. A. Trugman, J. X. Zhu, H. Ding, A. J. Taylor, D. A. Yarotski, and R. P. Prasankumar, Phys. Rev. B **92**, 161104(R) (2015).
[12] Y. Ishida, H. Masuda, H. Sakai, S. Ishiwata, and S. Shin, Phys. Rev. B **93**, 100302(R) (2016).
[13] E. Baek, T. Rim, J. Schutt, C. K. Baek, K. Kim, L. Baraban, and G. Cuniberti, Nano Lett. **17**, 6727 (2017).
[14] Y. Yang, X. Peng, H. S. Kim, T. Kim, S. Jeon, H. K. Kang, W. Choi, J. Song, Y. J. Doh, and D. Yu, Nano Lett. **15**, 5875 (2015).
[15] D. Khokhlov, L. Ryabova, A. Nicorici, V. Shklover, S. Ganichev, S. Danilov, and V. Bel'kov, Appl. Phys. Lett. **93**, 264103 (2008).
[16] B. A. Akimov and V. A. Bogoyavlenskiy, Phys. Rev. B **61**, 16045 (2000).
[17] R. Sreekumar, R. Jayakrishnan, C. Sudha Kartha, and K. P. Vijayakumar, J. Appl. Phys. **100**, 033707 (2006).
[18] V. K. Thorsmølle, R. D. Averitt, X. Chi, D. J. Hilton, D. L. Smith, A. P. Ramirez, and A. J. Taylor, Appl. Phys. Lett. **84**, 891 (2004).
[19] K. Lee, J. Li, L. Cheng, J. Wang, D. Kumar, Q. Wang, M. Chen, Y. Wu, G. Eda, E. E. M. Chia, H. Chang, and H. Yang, ACS Nano **13**, 9587 (2019).